\documentclass[twocolumn,showpacs,preprintnumbers,amsmath,amssymb,pra,epsf]{revtex4}
\usepackage{graphicx}
\usepackage{epsf,dsfont,hyperref,url,float}

\begin{document}

\title{Doves and hawks in economics revisited \\
{\it An evolutionary quantum game theory-based analysis of financial crises}}

\author{Matthias Hanauske$^a$}
\author{Jennifer Kunz$^b$}
\author{Steffen Bernius$^a$}
\author{Wolfgang K\"onig$^c$}
\affiliation{
$^a$Institute of Information Systems, $^b$Chair of Controlling \& Auditing, $^c$House of Finance\\
Goethe-University, Gr\"uneburgplatz 1, 60323 Frankfurt$/$Main
}

\date{\today}

\begin{abstract}  
The last financial and economic crisis demonstrated the dysfunctional long-term effects of aggressive behaviour in financial markets. Yet, evolutionary game theory predicts that under the condition of strategic dependence a certain degree of aggressive behaviour remains within a given population of agents. However, as the consequences of the financial crisis exhibit, it would be desirable to change the ''rules of the game'' in a way that prevents the occurrence of any aggressive behaviour and thereby also the danger of market crashes. The paper picks up this aspect. Through the extension of the in literature well-known Hawk-Dove game by a quantum approach, we can show that dependent on entanglement, also evolutionary stable strategies can emerge, which are not predicted by classical evolutionary game theory and where the total economic population uses a non aggressive quantum strategy.
\end{abstract}

\maketitle

\section{Introduction}\label{sec:intro}
Economic developments often have been compared to biological evolutionary processes, as they converge to equilibria in an evolutionary manner (e.g. Hodgson, 1993; Dosi $\&$ Nelson, 1994; Dopfer, 2001 \cite{book-hodgson,dosi-94,book-dopfer}). Actually, the conceptual ideas behind evolutionary theory were borrowed from early economic works, especially Malthus (1798) \cite{book-malthus} (see e.g. Friedmann, 1998 \cite{friedman-98}). Due to inter alia the application of evolutionary game theory, whose origin lies in biology (Maynard Smith, 1972, 1982 \cite{smith-1972,book_smith}), evolutionary concepts came back into economics and organisational theory. One major topic in this evolutionary research field is the optimality of aggressive versus non aggressive or cooperative behaviour (see e.g. for the tension of cooperative and non-cooperative behaviour Axelrod, 1997 \cite{book-axelrod}). In an economic context the notion of aggressive behaviour can be translated to the short-term oriented maximisation of individual utility without looking after others, while cooperative behaviour comprises a more interactive and long-term oriented behaviour considering long-term, individual and/or group utility maximisation. Possible positive effects of the mentioned aggressive behaviour on economic welfare have been discussed since the earliest days of economics (Smith, 1776 \cite{book-smith1776}): The idea was that if each economic individual tries to maximise his/her utility without caring about other individuals, the whole welfare will also be maximal. 

One instrument to analyse the long-term effects of this assumption is evolutionary game theory. Analogous to classical game theory it introduces the concept of strategic dependence among agents in an economic context. In such a situation the expected utility of one agent depends on the decisions of other agents. Evolutionary game theory provides an equilibrium in which the ratio of aggressive to non aggressive agents is stable and that depends on the expected losses and gains of utility induced by the agents’ decisions. For example, if the expected losses are high for two meeting aggressive agents, most members of the economic population -- but not all of them --  will behave in a none-aggressive, cooperative way (Osborne $\&$ Rubenstein, 1994 \cite{book-osborne}). Hence, also in situations where severe losses are expected, if two aggressive agents meet, an economic population always will contain a certain degree of aggressive agents. 

In economic reality, exactly this aspect can be observed, for example in the recent financial crisis: Each participant of financial transactions knew that highly risky financial products would increase the risk of the whole market portfolio and thereby augment the probability of a market crash resulting in huge losses. Nevertheless, several participants continued selling and buying these products in order to maximise their own, short-term utility resulting from high selling premiums and investment returns. Hence, these individuals followed an aggressive strategy. However, as the occurrence of the financial crisis exhibited, this behaviour can result in severe problems for the whole economic population. So, the question rises, whether there is a possibility to change the rules of the game in a way that protects populations from these severe problems by inhibiting the occurrence of aggressive behaviour. 

To answer this question the classical concept of evolutionary game theory shall be extended by another game theoretical development that is currently discussed: quantum games. The discussion of quantum games started with the work of Meyer (1999) \cite{meyer-1999-82} and Eisert et al. (1999) \cite{eisert-1999-83}. Meyer analysed the ''penny flip'' game and showed, that a player who selects a quantum strategy always wins this game. Eisert et al. (1999) concentrated on the prisoner dilemma and demonstrated that the players of this game could escape this dilemma if the entanglement of the prisoners’ wave function is above a certain value. Since these leadoff articles several further applications of quantum games have been published. Marinatto $\&$ Weber (2000) \cite{marinatto-2000-272} applied quantum games to the ''battle of sexes'' showing that entangled strategies will lead to a unique solution of this game. Benjamin $\&$ Hayden (2001) \cite{benjamin-2001-64} amplified the quantum game approach to a situation of multiple players. Piotrowski $\&$ Sladkowsky (2002) \cite{piotrowski-2002-312} used quantum games to examine market behaviour. Hanauske et al. (2007) \cite{hanauske-2006} based the analysis of the open access publishing behaviour in different scientific communities on a quantum game approach.  

The combination of this quantum game approach and evolutionary game theory has been applied by \cite{esteban-2006,toor-2001,toor-2001a}. We add to this existing research a practical application of this type of game theory.
Our results show that dependent on entanglement, also evolutionary stable
strategies can emerge, which are not predicted by classical
evolutionary game theory: The analysis exhibits the existence of a new,
payoff dominant evolutionary stable strategy (ESS), where the whole
economic population uses the non aggressive quantum dove strategy.
We interpret entanglement in this context as the objective influence of socio-economic 
context factors, while the application of quantum strategies exhibits the degree to which 
decision makers incorporate these factors into their decisions.
This interpretation allows the derivation of consequences and shows the
linkage of our study to other game theoretical analyses that also
highlight the importance of the socio-economic context to the outcomes
of games. For example, Sally \cite{Sally.2001} discusses the notion of sympathy, a
feeling that occurs when players get to know each other and that can
lead to increasing cooperation in prisoners' dilemma games. 

The paper is structured as follows: We pick up the recent financial crisis as an example for the fruitful application of evolutionary quantum game theory. In order to do so, we have to select a group of participants in the financial transactions that finally lead to the crisis. We have chosen the group of inventors and sellers of the highly risky financial products. Their behaviour can be interpreted as the in theory well known Hawk-Dove game (Maynard Smith 1982, 1986 \cite{book_smith,smith-1986}). Hence, in section \ref{jenny2} we develop a model that is based on this game type and comprises the relevant parts of the behaviour of these constructors and sellers to mirror the starting conditions of the financial crisis. In section \ref{classical} we transfer this model into a classical evolutionary game. Section \ref{quantum} is dedicated to the quantum version of this game, while section \ref{qeg} comprises the evolutionary quantum version. In section \ref{conclusions} we draw some conclusions from our findings. The paper closes with a summary in section \ref{sec:sum}. 

\section{The financial crisis as Hawk-Dove game}\label{jenny2}
Financial crises in general and the last one especially, have their origin in highly speculative behaviour of market participants. In our analysis we focus on a specific population of market participants, who had a great part in the last crisis: Constructors and sellers of investment papers with different degrees of risk. They played an important role in the last crisis as follows:\\
This crisis grounded especially on the housing market in the United States. Based on the idea of continuously increasing prices for real estates, loans were also provided to borrowers, who actually could not afford buying a house. But under the premise of increasing house values, providing loans to these people seemed to be rational as they were backed by increasingly valuable real estates. Yet, these loans did not remain with the lending credit institutes but they were bundled to portfolios together with loans of higher solvency. These portfolios then were sold to other banks as investment products. The idea behind these products is to spread risk among banks. Moreover, papers of higher risk also promise a higher return, which makes them attractive for speculative purposes. The buying banks often unbundled the loans and bundled them together with other loans to sell again parts of these newly created portfolios. These processes were repeated several times. So finally, the loans were scattered around the world. However, after the house prices started falling, the bad loans became obvious in these portfolios and caused losses. But, as the loans were scattered around the world, nobody really knew where which risk still remained and which bank would suffer next from a financial disorder. As a result of this, banks stopped providing credit to each other in order to prevent credit defaults. This trust crisis actually lead to the severe economic problems, as not only banks but also other firms got problems to receive credits for the continuation of their business. 

Hence, one major driver of the crisis was the mentioned speculative investment products. The described portfolios had a considerable degree of complexity. In combination with the continuously spreading of risks among the same investors it was only a matter of time that the crisis had to start. However, although this was foreseeable dealing with this investment products continued. This scenario can be transferred to a model usable for evolutionary game theory as follows:\\
In line with the classical Hawk-Dove model two types of agents shall be considered: Doves follow a non aggressive strategy. Transferred to the financial situation they are investment bankers who construct investment products of rather low risk and moderate expected return. These products lead to a moderate premium to the seller but have no negative long-term impact on the total market risk. Additionally, when selling their products to investors, doves remain with their contract conditions and do not try to make a deal by all means, e.g. promising unrealistic returns or omitting to point out severe risk factors of the investment product. In contrast, hawks follow an aggressive strategy. They represent those investment bankers, who are specialised on highly risky products with high expected returns. They also act aggressive to sell their products, which might end up in investment constructs that contain a destabilising potential to the financial market. Both types of agents ''fight'' for a pool of risk-neutral investors. For simplification reasons, we assume that always only two agents fight for one investor, where both agents can be doves, or hawks, or one is a dove and one is a hawk.

If a dove and a hawk fight for one investor, the hawk will win, as he/she can offer a product with a higher expected return. If two doves meet, the investor will spread the investment equally, as it is assumed that both offer him/her the same conditions. If two hawks meet, the investor will also spread the investment equally, as again it is assumed that both hawks provide the same investment product. However, the payoffs of the players are quite different in all three cases and contain two parts.

The first part is the selling premium. This premium depends on the expected return of the sold investment product. In the first case, the dove gets nothing, as it cannot sell any product, while the hawk receives a high premium $p_h$. In the second case, both doves get half of the moderate premium $p_m$, as the investment sum is split up between both. In the last case, both hawks receive half of the high premium, as again the investment is split up. 

The second part comprises a discount resulting from the fight of two players for one investor. In the first case, an aggressive and a non aggressive investment banker meet. Here, no fight will take place, as the non aggressive banker remains with his/her conditions and the investor prefers the product with the higher expected return. Hence, the aggressive banker has no reason to start any fight, since he/she can sell his/her product. Regarding the second case, again no fighting will be observed, as both bankers stay with their conditions and the investor just splits up the investment sum. Consequently, in the first and the second cases, no discount has to be considered. However, if two aggressive bankers meet, they will try to get the whole investment sum and start fighting for it. On the one hand, this can result in a lowering of selling prices. On the other hand, this ends in the construction of products which offer an even higher expected return but bear very high, partly hidden risks. These effects are totalled in a discount parameter $d$. Hence, both aggressive bankers receive half of the high premium minus this discount. The discount factor is an indicator for the degree of aggressiveness of the hawks and at the same time for the danger of the products resulting from the meeting of two hawks to cause a future crash due to hidden risk. Table \ref{tab:PayOff_hawkdove} summarises the payoff matrix.

\begin{table}[H]
\centerline{
\begin{tabular}{|r|c c|}
        \hline
        A$\setminus$B& Hawk & Dove
        \\ \hline
        Hawk & ($\frac{p_h - d}{2}$,$\frac{p_h - d}{2}$) & ($p_h$,0)
        \\
        Dove & (0,$p_h$) & ($\frac{p_m}{2}$,$\frac{p_m}{2}$)\\\hline
\end{tabular}}
\caption[caption]{Payoff matrix for investment bankers $A$ and $B$ within the Hawk-Dove game. The parameters are defined as follows: $p_h$: high selling premium, $d$: disutility resulting from fighting and $p_m$: moderate selling premium.}
\label{tab:PayOff_hawkdove}
\end{table}
To assure the payoff matrix to have the formal structure of a Hawk-Dove game the parameters of Table \ref{tab:PayOff_hawkdove} should fulfil the inequation $p_h > p_m > 0 > \frac{p_h - d}{2}$, which means that the disutility $d$ should be higher than the high selling premium $p_h$.

\section{The classical evolutionary game of doves and hawks}\label{classical}
This section is dedicated to the introduction of the necessary definitions and fundamental basics of an evolutionary game. In the following the presentation is constrained to describe the mixed enlargement of a symmetric two person, n--strategy game $\Gamma$ (for details see \cite{book_schlee_gametheory}):

\begin{eqnarray}
\Gamma := \left(  \left\{ A, B \right\}, {\bf \cal S} \times {\bf \cal S}, \hat{\bf {\cal \$}}  \right) && \mbox{: 2-person game}\nonumber\\
s = \left( s_1, s_2, ..., s_n \right) \in {\bf \cal S}  && \mbox{: Mixed strategy}\nonumber\\
\hat{\bf {\cal \$}} = \left(
\begin{array}[c]{cccc}
\$_{11}&\$_{12}& ... &\$_{1n}\\
\$_{21}&\$_{22}& ... &\$_{2n}\\
... & ... & ... & ...\\
\$_{n1}&\$_{n2}& ... &\$_{nn}\\
\end{array}
\right) && \mbox{: Payoff matrix}
\label{2*N-Spiel}
\end{eqnarray}

To describe the time evolution of the repeated version of the game $\Gamma$, replicator dynamics were developed. Replicator dynamics, formulated within a system of differential equations, defines in which way the population vector $\vec{x}:=(x_1, x_2, ..., x_n)$ evolves in time. Each component $x_i=x_i(t)$ ($i=1, 2, ..., n$) describes the time evolution of the fraction of different player types $i$ in the whole population, where a type-$i$ player is understood as an actor playing strategy $s_i$. The population vector $\vec{x}$ has to fulfil the normalising conditions of a unity vector 
\begin{equation}
x_i(t) \geq  0 \quad \forall \, i=1, 2, ..., n \, , \,\, t \in \mathds{R} \, \mbox{and} \, 
\sum_{i=1}^n x_i(t) = 1  \label{Glei:population1}.
\end{equation}

The following first order system of differential equations of the population vector $\vec{x}(t)=(x_1(t), x_2(t), ..., x_n(t))$ is known as replicator dynamics (see \cite{miekisz-2007,book_schlee_gametheory})
\begin{equation}
\frac{d x_i(t)}{dt} =  x_i(t) \left[ \underbrace{\sum_{l=1}^n \$_{il} \, x_l(t)}_{:=f_i(t)} - 
\underbrace{\sum_{l=1}^n \sum_{k=1}^n \$_{kl} \, x_k(t) \, x_l(t)}_{:=\bar{f}(t)} \right] \label{Glei:Repro}
\end{equation}
where $f_i(t)$ is the fitness of type $i$ and $\bar{f}(t)=\sum_{i=1}^n f_i(t)$ is the avarage fitness of the whole population. 

In the following the formal description is restricted to only two strategies ($i= 1,2 \, \hat{=} \, H,D$). Because of condition \ref{Glei:population1}, the population vector $\vec{x}(t)=(x_1(t), x_2(t))$ can be reduced to only one independent component ($x(t):=x_1(t)$, and $x_2(t)=1-x(t)$) and equation \ref{Glei:Repro} simplifies as follows: 

\begin{equation}
\frac{d x}{dt} = x \left[ (\$_{11} -  \$_{21}) (x-x^2) + (\$_{22} - \$_{12}) (1-2x+x^2) \right] \nonumber
\end{equation}

Inserting the parameters of the Hawk-Dove payoff matrix (see Table \ref{tab:PayOff_hawkdove}) gives the following differential equation: 

\begin{eqnarray}
\frac{d x}{dt} &=& \frac{1}{2}\left( p_h - p_m +d \right) \, x^3 + \left( p_m - \frac{3}{2} p_h - \frac{1}{2} d \right) \, x^2 + \nonumber
\\
&& + \left( p_h - \frac{1}{2} p_m \right) \, x \label{Glei:Repro2}
\end{eqnarray}

To show the consequences of equation \ref{Glei:Repro2} and to discuss and illustrate the main properties of the underlying Hawk-Dove game the payoff parameters of Table \ref{tab:PayOff_hawkdove} have been set to the following three different parameter sets. 

\begin{table}[H]
\centerline{
\begin{tabular}{|c|c|c|c|c|}
        \hline
        Parameter  & Risk of & &  &
        \\ 
        setting & destabilisation & $d$ & $p_h$ & $p_m$
        \\ \hline
        P1 & LOW & 6 & 5 & 3
        \\ \hline
        P2 & MEDIUM & 10 & 5 & 3
        \\ \hline
        P3 & HIGH & 20 & 5 & 3
        \\\hline
\end{tabular}}
\caption[caption]{Parameters of the three different sets of the underlying payoff matrix used to model the investment market of the Hawk-Dove game.}
\label{tab:Parameters_hawkdove}
\end{table}

Within the parameter sets the high and low selling premiums are fixed ($p_h=5$ and $p_m=3$), whereas the destabilising factor $d$ is varied. In parameter set $P1$ the risk of destabilisation is only a little bit higher ($d=6$) than the high selling premium, in parameter set $P2$ a medium value of the destabilising factor $d$ that results from fighting was used ($d=10$), and in set $P3$ the parameter $d$ was chosen to a quite high value ($d=20$). 

\begin{figure}[b]
\vspace*{-2.0cm}
\centerline{
\includegraphics[width=3.6in]{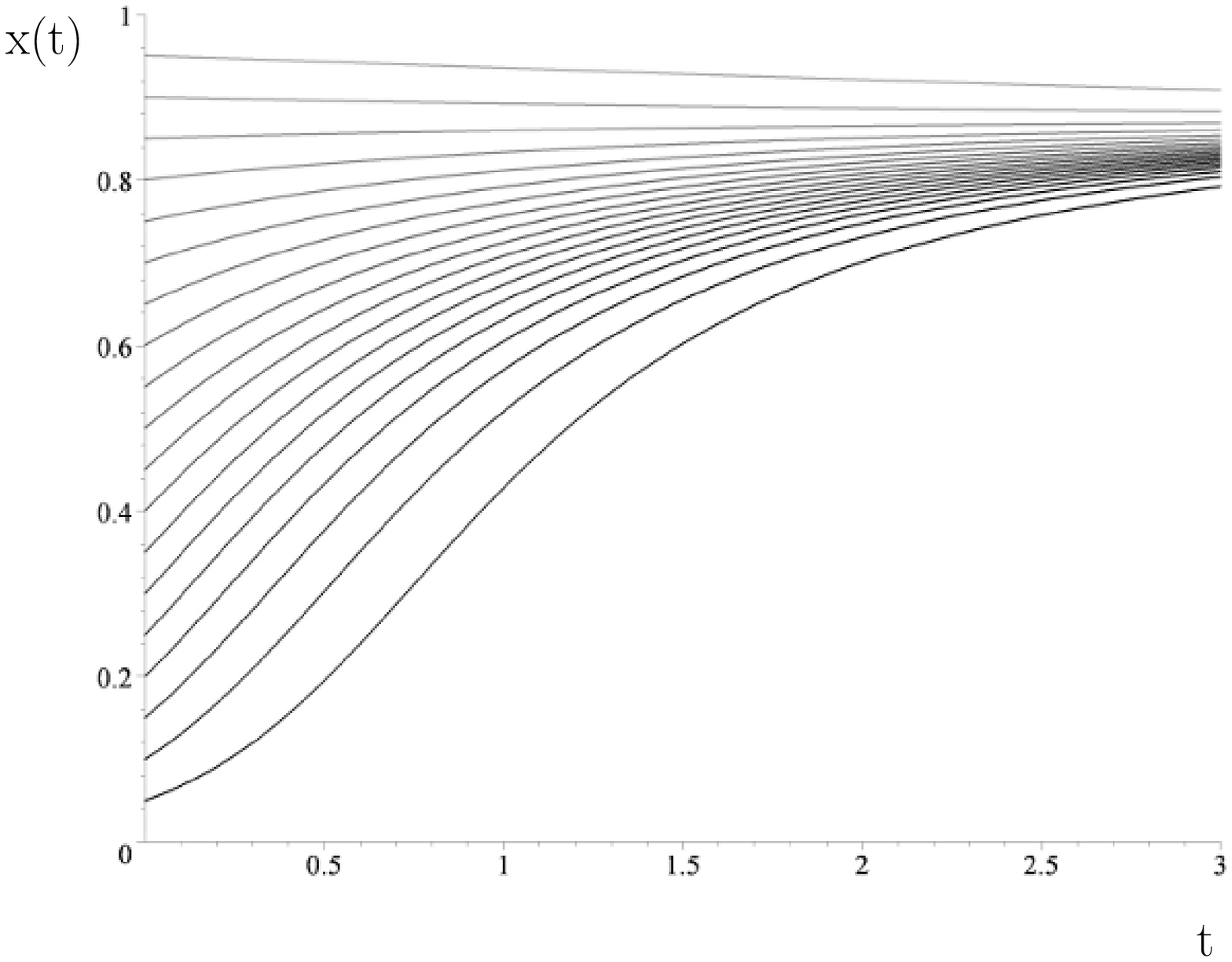}
}
\vspace*{-5.0cm}
\caption{Fraction of hawks $x$ as a function of time $t$ for different starting values $x(t=0)$. Results were calculated using the parameter set $P1$ (low risk investment market).}
\label{fig:lowrisk}
\end{figure}

\begin{figure}[h]
\vspace*{-2.0cm}
\centerline{
\includegraphics[width=3.6in]{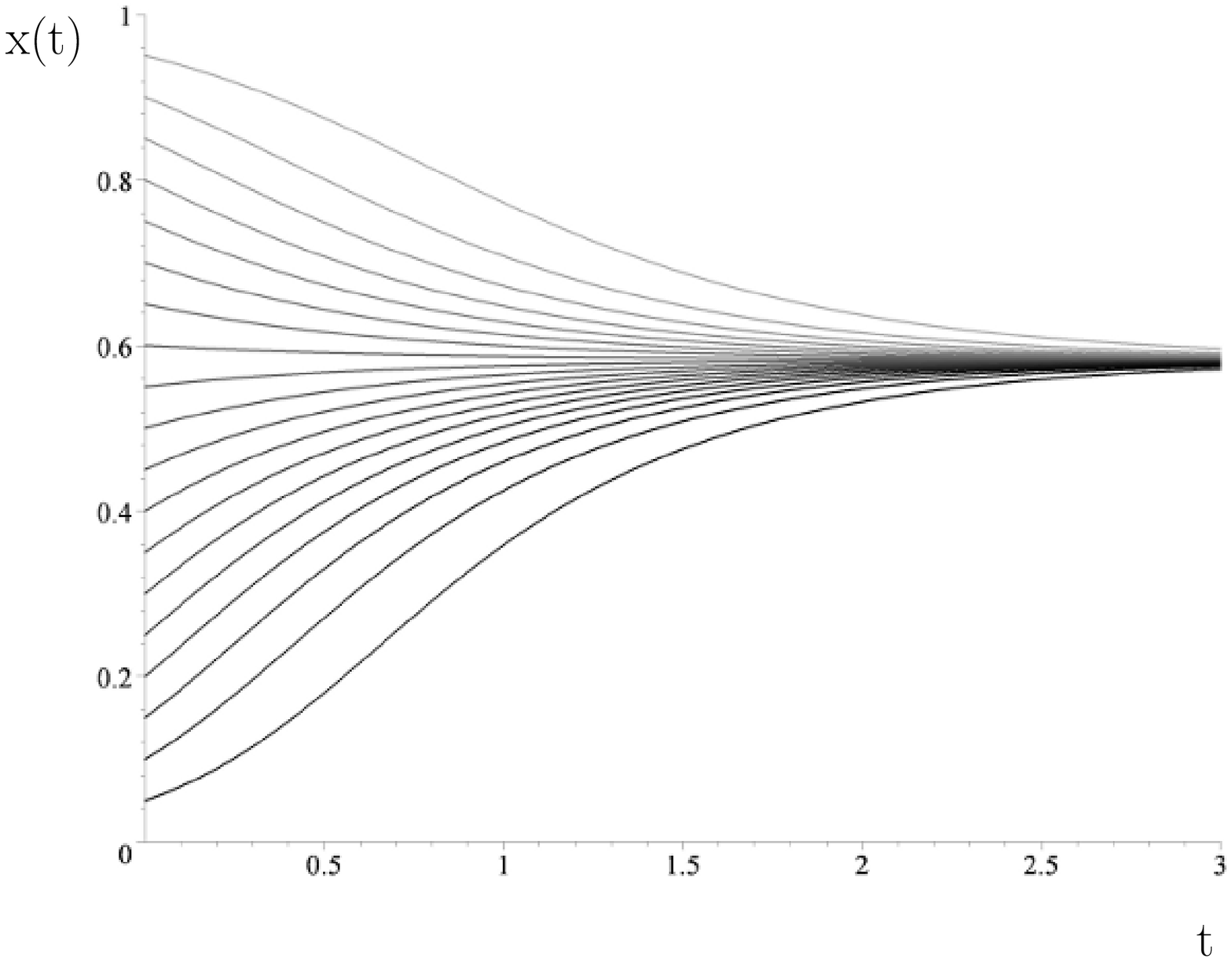}
}
\vspace*{-5.0cm}
\caption{Description like in Figure \ref{fig:lowrisk}. Results were calculated using the parameter set $P2$ (medium risk investment market).}
\label{fig:mediumrisk}
\end{figure}

\begin{figure}[h]
\vspace*{-2.0cm}
\centerline{
\includegraphics[width=3.6in]{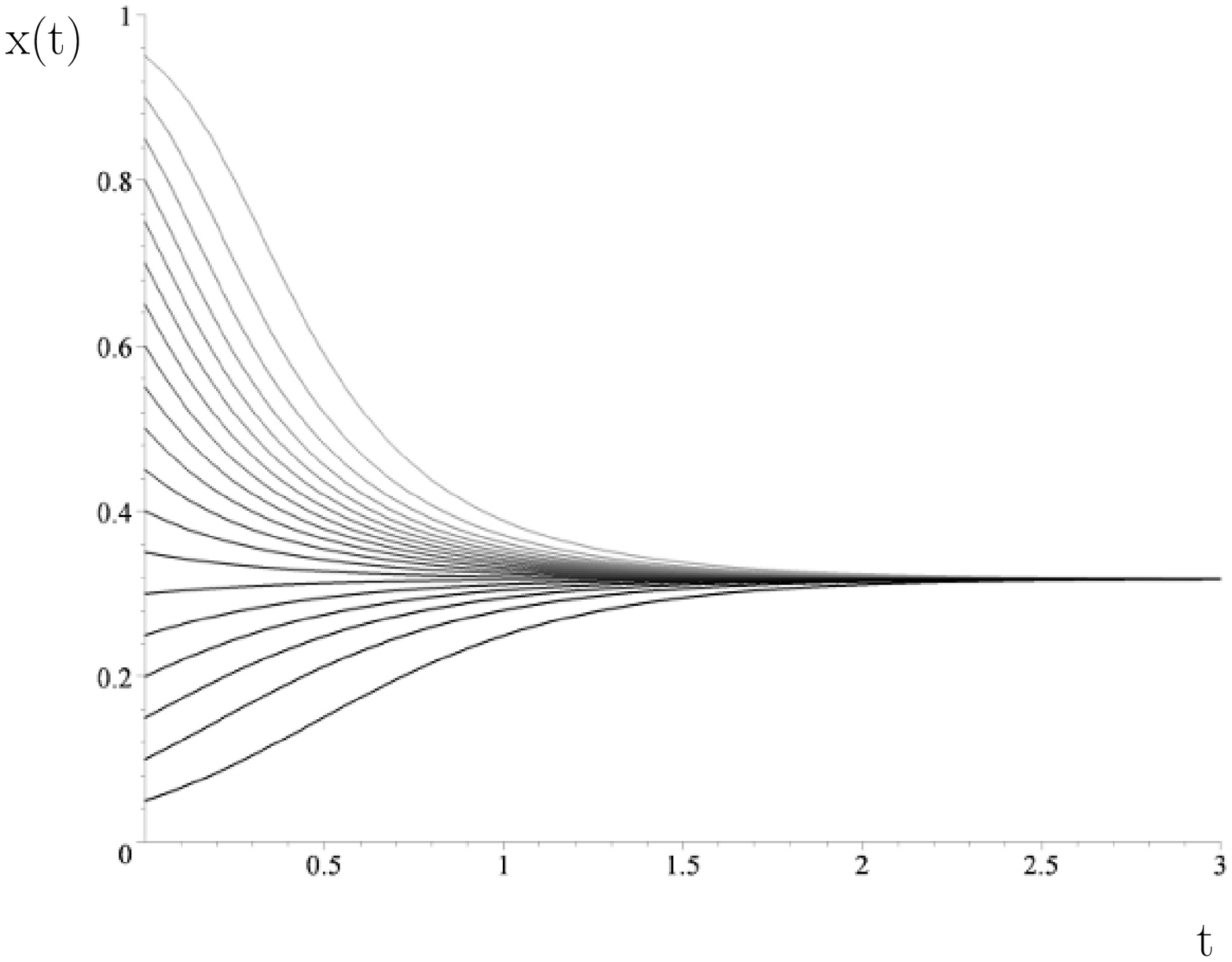}
}
\vspace*{-5.0cm}
\caption{Description like in Figure \ref{fig:lowrisk}. Results were calculated using the parameter set $P3$ (high risk investment market).}
\label{fig:highrisk}
\end{figure}

The evolution of the fraction of hawks $x(t)$ within the hawk-dove population is displayed in Figures \ref{fig:lowrisk}, \ref{fig:mediumrisk} and \ref{fig:highrisk}. Figure \ref{fig:lowrisk} shows $x(t)$ as a function of time for the parameter set $P1$, in which the different curves where calculated using various different starting values of the fraction of hawks at time zero ($x(0)=\frac{1}{20}, \frac{2}{20}, ..., \frac{19}{20}$). The Figure shows clearly that all population curves converge to one limit value $x_L:=x(t \rightarrow \infty)$. Within parameter set $P1$ the fraction of hawks ends after some time always at $x_L=0.86$, which means that the population of hawks and doves will be stable if it consists of $86 \%$ hawks and $14 \%$ doves. Parameter set $P1$ corresponds to a situation where the risk of a future crash of the whole investment market is expected to be quite low. Within such a situation the theory predicts that the relative number of investment bankers selling highly risky products (hawk strategy) is quite high ($86 \%$) and as a result the fraction of sellers offering products with moderate returns and a rather low risk is quite low ($14 \%$).  

Within parameter set $P2$ the underlying investment market has a medium crashing risk. Figure \ref{fig:mediumrisk} shows that for such a market the stable fraction of hawks (investment bankers selling highly risky products) has decreased ($x_L=0.56$). 

Figure \ref{fig:highrisk} shows the situation where aggressive behaviour will lead to an unstable market, in which it is very like that a future crash will occur.  
The players within such a highly risky market choose mainly a non risky dove strategy ($x_L=0.34$), but still $34 \%$ of the investment bankers sell highly risky products.

To understand the simulated results more formally, the concept of evolutionary stable strategies is briefly explained in the following.\\ 

\fbox{
\begin{minipage}[H]{7cm}
Taking a general symmetric 2-player game $\Gamma$ with a payoff matrix $\hat{\bf {\cal \$}}$. A strategy $s^* \in {\bf \cal S}$ is defined as an evolutionary stable strategy (ESS) if \cite{book_schlee_gametheory} \\
a) $(s^*,s^*)$ is a Nash equilibrium of the game\\
b) $\$(s,s) \le \$(s^*,s) \,\,\, \forall \,\, s \in r(s^*) \, , \,\,s \neq s^*$
\end{minipage}
}
\vspace{0.3cm}

$r(s^*)$ is the best response function to the strategy $s^*$ and $\$(s,s)$ describes the extended, mixed strategy payoff function. An evolutionary stable strategy $s^*$ therefore needs to be a symmetric Nash equilibrium of the game and additionally the inequation b) should be fulfilled for any strategy $s$ belonging to the set of best responses to $s^*$ ($s \in r(s^*)$). To illustrate this definition we restrict the number of pure strategies to $n=2$ and use the payoff matrix of Table \ref{tab:PayOff_hawkdove}. $x:=s_1^A$ denotes the probability of player A playing the aggressive strategy hawk, while $y:=s_1^B$ defines the probability of player B playing strategy hawk. The mixed strategy payoff function has therefore the following structure:
\begin{eqnarray}
\$(x,y) &=& \$_{11} \,\, x \, y + \$_{12} \,\, x \, (1-y) + \nonumber \\
&& + \$_{21} \,\, (1 - x) \, y + \$_{22} \,\, (1 - x) \, (1-y) \\
&=& \frac{p_h - d}{2} \, x \, y + p_h \, x \, (1-y) + \frac{p_m}{2} \, (1 - x) \, (1-y) \nonumber 
\label{Glei:Payoffmsp}
\end{eqnarray}
Because of the symmetry of the game, the payoff of player A ($\$^A(x,y) = \$(x,y)$) and the payoff of player B are equal after variable transformation ($x \rightarrow y, y \rightarrow x$).

\begin{equation}
\$^A(x,y) = \$(x,y) \quad \hbox{and} \quad \$^B(x,y) = \$(y,x) \nonumber
\end{equation}

The two necessary conditions to prove the existence of a Nash equilibrium $(x^*,y^*)$ in a 2-person 2-strategy game reduce therefore to one single condition \cite{book_schlee_gametheory}

\begin{eqnarray}
\$^A(x^*,y^*) &\geq& \$^A(x,y^*) \quad \forall \,\,\, x \in [0,1] \nonumber\\
\$^B(x^*,y^*) &\geq& \$^B(x^*,y) \quad \forall \,\,\, y \in [0,1] \nonumber\\
\Rightarrow \$(x^*,y^*) &\geq& \$(x,y^*) \quad \forall \,\,\, x \in [0,1] 
\label{Glei:Nashequsym}
\end{eqnarray}

\begin{figure}[h]
\vspace*{-1.5cm}
\centerline{
\includegraphics[width=4.0in]{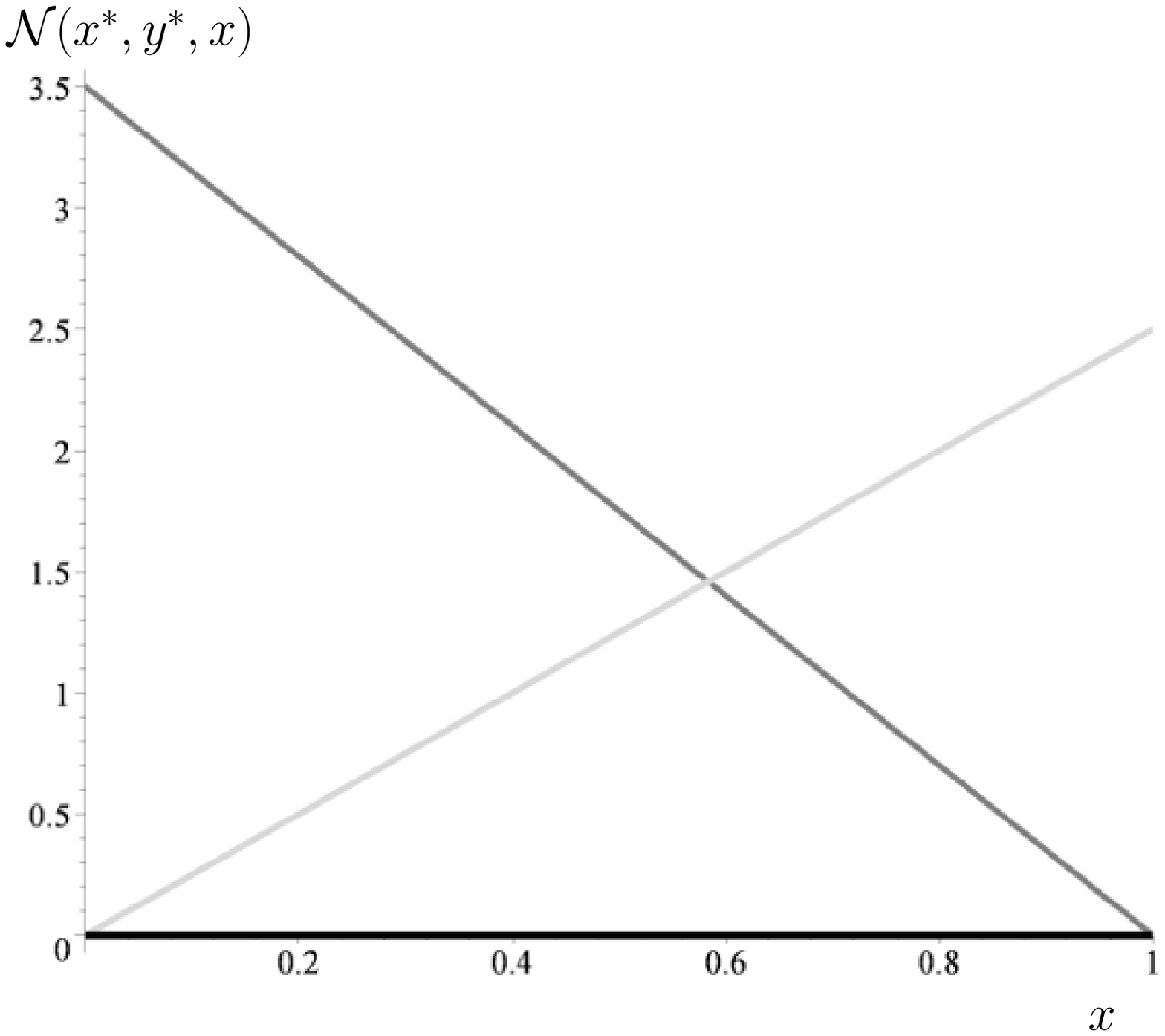}
}
\vspace*{-5.8cm}
\caption{${\cal N}(x^*,y^*,x)$ for the three Nash equilibria as a function of $x$ within the middle risk parameter setting $P2$. The dark grey line corresponds to the Nash equilibrium $(x^*,y^*)=(1,0)$, the light grey line to $(x^*,y^*)=(0,1)$ and the black line (${\cal N} \equiv 0$) to the mixed strategy Nash equilibrium $(x^*,y^*)=(\frac{p_m - 2 p_h}{p_m - p_h -d},\frac{p_m - 2 p_h}{p_m - p_h -d})$}
\label{fig:nashproof}
\end{figure}

The game has three Nash equilibria. Two are non symmetric pure Nash equilibria ($(x=1,y=0) \hat{=} (H,D)$ and $(x=0,y=1) \hat{=} (D,H)$) and one is a symmetric, mixed strategy Nash equilibrium $(x=\frac{p_m - 2 p_h}{p_m - p_h -d},y=\frac{p_m - 2 p_h}{p_m - p_h -d})$. Definition \ref{Glei:Nashequsym} requires that in every Nash equilibrium the function ${\cal N} := \$(x^*,y^*) - \$(x,y^*)$ needs to be positive for all $x \in [0,1]$. Figure \ref{fig:nashproof} shows ${\cal N}$ for the three Nash equilibria within a middle risk szenario (parameter set $P2$) and proves their existence, as all values are non negative. 

To prove the existence of an evolutionary stable strategy, condition b) has to be fulfilled additionally for the mixed strategy Nash equilibrium. The best response of player A to the strategy $y^*=\frac{p_m - 2 p_h}{p_m - p_h -d}$ is the set of all strategies $x \in [0,1]$, because $\$(x,y^*) = -\frac{(p_h -d) p_m}{2 (p_h + d - p_m)}$ is independent of x. Condition b) therefore has to be checked for all $x \in [0,1] \setminus \{ \frac{p_m - 2 p_h}{p_m - p_h -d} \}$. $x^*$ is an ESS if the function ${\cal G}(x^*,x)$ fulfils the following condition:
\begin{eqnarray}
&{\cal G} := \$(x^*,x) - \$(x,x) \ge 0 &\\
&\forall \, x \in [0,1] \setminus \{ x^*=\frac{p_m - 2 p_h}{p_m - p_h -d} \}& \nonumber
\end{eqnarray}

\begin{figure}[h]
\vspace*{-1.0cm}
\centerline{
\includegraphics[width=4.0in]{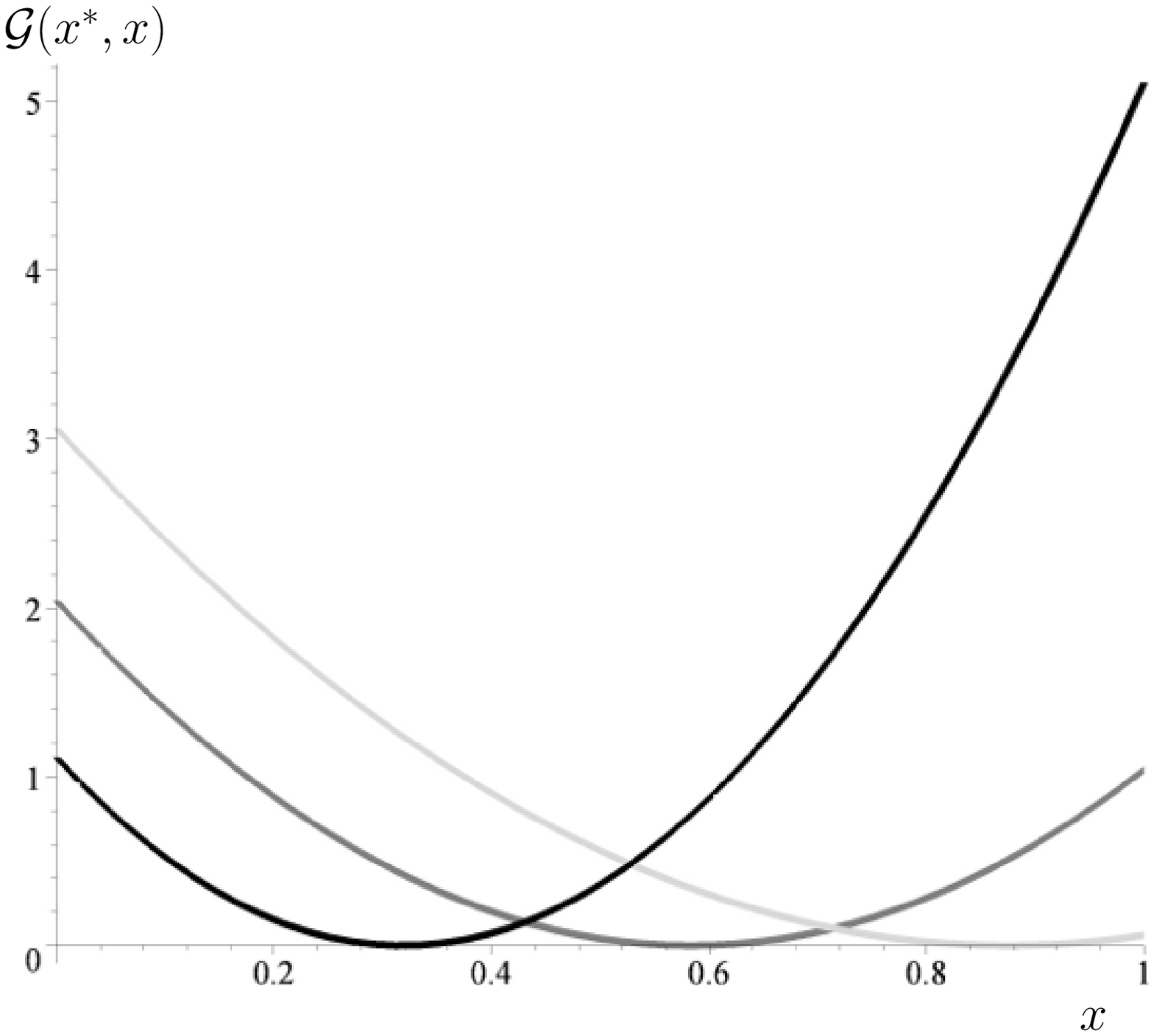}
}
\vspace*{-5.8cm}
\caption{${\cal G}(x^*,x)$ for the three different parameter settings $P1$ ($x^*=\frac{7}{8}$, light grey curve), $P2$ ($x^*=\frac{7}{12}$, dark grey curve) and $P3$ ($x^*=\frac{7}{22}$, black curve).}
\label{fig:ESSproof}
\end{figure}

Figure \ref{fig:ESSproof} shows ${\cal G}$ for the symmetric Nash equilibria $x^*=\frac{7}{2+d}$ ($p_h=5, p_m=3$) of the three different payoff parameter settings ($d=6, 10, 20$). The null of the three curves corresponds to the evolutionary stable fraction of hawk strategies within the low ($d=6$), middle ($d=10$) and high ($d=20$) risk settings. As the destabilisation risk $d$ increases, the ESS $x^*=\frac{7}{2+d}$ (the fraction of hawks) decreases. 

In sum, the results of the previous analysis based on evolutionary game theory suggest that dependent on the destabilisation factor the degree of aggressive agents varies, but even in case of highly risky markets aggressive behaviour will not vanish completely. This is exactly, what could be observed previously to the financial crisis: Although the risk of destabilisation in the investment market was obviously increasing for the last few years, the behaviour of some aggressive investment bankers did not change. However, instead of ending in a stable state, finally the market crashed and almost all aggressive agents disappeared from the population. This could have been prevented, if any aggressive behaviour were inhibited completely. 

It will be shown in this article that a quantum game theoretical formulation of the Hawk-Dove game is able to induce exactly this result: within the subset of quantum dove strategies, it will be shown that if the strategy of all investment bankers is entangled above a certain value, a new evolutionary stable quantum strategy is possible, leading to an observed banker population offering solely non aggressive investment products. Compared to the classical mixed strategy Nash equilibrium of the game, the new evolutionary stable quantum strategy is payoff dominant, if the strength of entanglement is above a certain value. To describe these phenomena in a more detailed way, the quantum game theory of doves and hawks is addressed within the following two sections.

\section{The quantum game of doves and hawks}\label{quantum}
In quantum game theory, the measurable pure classical strategies (H and D) correspond to the orthonormal unit basis vectors $\left| H \right>$ and $\left| D \right>$ of the two dimensional complex space $\mathds{C}^2$, the so called Hilbert space ${\cal{H}}_i$ of player $i$ ($i=A,B$). A quantum strategy of a player i is represented as a general unit vector $\left| \psi \right>_i$ in his strategic Hilbert space ${\cal{H}}_i$. The whole quantum strategy space $\cal{H}$ is constructed with the use of the direct tensor product of the individual Hilbert spaces: ${\cal{H}}:={\cal{H}}_A \otimes {\cal{H}}_B$. The main difference between classical and quantum game theory is that in the Hilbert space ${\cal{H}}$ correlations between the players' individual quantum strategies are allowed, if the two quantum strategies $\left| \psi \right>_{\!A} \in {\cal{H}}_A$ and $\left| \psi \right>_{\!B}  \in {\cal{H}}_B$ are entangled. The overall state of the system we are looking at is described as a 2-player quantum state $\left| \Psi \right> \in {\cal{H}}$. We define the four basis vectors of the Hilbert space ${\cal{H}}$ as the classical game outcomes ($\left| DD \right>:=(1,0,0,0)$, $\left| DH \right>:=(0,-1,0,0)$, $\left| HD \right>:=(0,0,-1,0)$ and $\left| HH \right>:=(0,0,0,1)$). 

The setup of the quantum game begins with the choice of the initial state $\left| \Psi_0 \right>$. We assume that both players are in the state $\left| D \right>$. The initial state of the two players is given by 
\begin{equation}
\left| \Psi_0 \right> \,=\, \widehat{\mathcal{J}} \left| DD \right> \,=\,
\left( \begin{array}{c}
 \cos\left( \frac{\gamma}{2} \right) \\ \\ 0 \\ \\ 0 \\ \\ i\sin\left( \frac{\gamma}{2}\right) \\
 \end{array} \right)  \quad ,
\end{equation}
where the unitary operator $\hat{\cal{J}}$ (see equation \ref{eq.J}) is responsible for the possible entanglement of the 2-player system. The players' quantum decision (quantum strategy) is formulated with the use of a two parameter set of unitary $2\times2$ matrices: 
\begin{eqnarray}
&\widehat{\mathcal{U}}(\theta,\varphi) :=
\left(
\begin{array}[c]{cc}
e^{i\,\varphi} \, \hbox{cos}(\frac{\theta}{2})&\hbox{sin}(\frac{\theta}{2})\\
-\hbox{sin}(\frac{\theta}{2})&e^{-i\,\varphi} \, \hbox{cos}(\frac{\theta}{2})
\end{array}
\right)&\\
&
\forall \quad \theta \in{} [0,\pi] \,\, \wedge \,\, \varphi \in{} [0,\frac{\pi}{2}] &\quad .\nonumber
\end{eqnarray}
By arranging the parameters $\theta$ and $\varphi$, a player chooses his quantum strategy. The classical strategy D (Dove) is selected by appointing $\theta=0$ and $\varphi=0$ :
\begin{equation}
\widehat{D}:=
\hat{{\cal{U}}}(0,0) =
\left(
\begin{array}[c]{cc}
1&0\\
0&1
\end{array}
\right)\quad,
\end{equation}
whereas the strategy H (Hawk) is selected by choosing $\theta=\pi$ and $\varphi=0$ :
\begin{equation}
\widehat{H}:=
\hat{{\cal{U}}}(\pi,0) =
\left(
\begin{array}[c]{cc}
0&1\\
-1&0
\end{array}
\right)\quad.
\end{equation}
In addition, the quantum strategy $\widehat{Q}$ is given by
\begin{equation}
\widehat{Q}:=
\hat{{\cal{U}}}(0,\pi/2) =
\left(
\begin{array}[c]{cc}
i&0\\
0&-i
\end{array}
\right)\quad.
\end{equation}

After the two players have chosen their individual quantum strategies ($\hat{{\cal{U}}}_A:=\hat{{\cal{U}}}(\theta_A,\varphi_A)$ and $\hat{{\cal{U}}}_B:=\hat{{\cal{U}}}(\theta_B,\varphi_B)$) the disentangling operator $\widehat{\mathcal{J}}^\dagger$ is acting to prepare the measurement of the players' state. The entangling and disentangling operator ($\widehat{\mathcal{J}}, \widehat{\mathcal{J}}^\dagger$; with $\hat{\cal{J}}\equiv\hat{\cal{J}}^\dagger$) depends on one additional single parameter $\gamma$ which measures the strength of the entanglement of the system:
\begin{equation}
\widehat{\mathcal{J}} := e^{i \, \frac{\gamma}{2} (\widehat{D} \otimes \, \widehat{D})} \,\, , \quad \gamma \in{} [0,\frac{\pi}{2}] \quad \label{eq.J}.
\end{equation}
In the used representation, the entangling operator $\widehat{\mathcal{J}}$ has the following explicit structure:  
\begin{equation}
\widehat{\mathcal{J}} := \left( \begin{array}{cccc}
 \cos\left( \frac{\gamma}{2} \right) & 0 & 0 & i\sin\left( \frac{\gamma}{2} \right) \\
  & & & \\
 0 & \cos\left( \frac{\gamma}{2}\right) & -i\sin\left( \frac{\gamma}{2}\right) & 0 \\
  & & & \\
 0 & -i\sin\left( \frac{\gamma}{2}\right) & \cos\left( \frac{\gamma}{2}\right) & 0 \\
  & & & \\
 i\sin\left( \frac{\gamma}{2}\right) & 0 & 0 & \cos\left( \frac{\gamma}{2}\right) \\
 \end{array} \right) \, .
\end{equation}

Finally, the state prior to detection can therefore be formulated as follows:
\begin{equation}
\left| \Psi_f \right> = \hat{\cal{J}}^\dagger \left( \hat{\cal{U}}_A \otimes \hat{\cal{U}}_B \right) \hat{\cal{J}}\, \left| DD \right> \quad .
\end{equation}
The expected payoff within a quantum version of a general 2-player game depends on the payoff matrix (see Table \ref{tab:PayOff_hawkdove}) and on the joint probability to observe the four observable outcomes $P_{\mbox{\small HH}}, P_{\mbox{\small HD}}, P_{\mbox{\small DH}}$ and $P_{\mbox{\small DD}}$ of the game  
\begin{eqnarray}
&\$_A&= \$_{11}\,P_{\mbox{\small  HH}} + \$_{12}\,P_{\mbox{\small HD}} + \$_{21}\,P_{\mbox{\small DH}} + \$_{22}\,P_{\mbox{\small DD}} \nonumber\\
&\$_B&= \$_{11}\,P_{\mbox{\small HH}} + \$_{21}\,P_{\mbox{\small HD}} + \$_{12}\,P_{\mbox{\small DH}} + \$_{22}\,P_{\mbox{\small DD}} \nonumber\\
&\mbox{with:}& P_{\sigma \sigma^{,}}=\left| \, \left< \sigma\sigma^{,} | \Psi_f \right> \, \right|^2 \,\, , \quad \sigma,\sigma^{,}=\left\{ H,D \right\} \quad . 
\end{eqnarray}

It should be pointed out here, that an entangled 2-player quantum state does not mean at all that the persons themselves (or even the players' brains) are entangled. The process of quantum decoherence, with its quantum to classical transition, forbid such macroscopic entangled systems established from microscopic quantum particles \cite{schlosshauer-2004,book_kiefer}. However, peoples' cogitations, represented as quantum strategies, could be associated within an abstract space. Although no measurable accord is present between the players' strategy choices, the imaginary parts of their strategy wave functions might interact, if their individual states are entangled. In the context of the financial investment market, this quantum phenomenon might possibly be interpreted as a conjoint, psychological contract between the investment bankers aligning their strategies and resulting from the impact of socio-economic context factors. Such an alignment is now formulated as the appearance of a strongly entangled strategy effectuating the players to act more like a collective state.

To visualise the payoffs in a three dimensional diagram it is necessary to reduce the set of parameters in the final state: $\left| \Psi_f \right>=\left| \Psi_f(\theta_A,\varphi_A,\theta_B,\varphi_B) \right> \rightarrow \left| \Psi_f(\tau_A,\tau_B) \right>$. Within the following diagrams we have used the same specific parameterisation as Eisert et al. \cite{eisert-1999-83}, where the two strategy angles $\theta$ and $\varphi$ depend only on a single parameter $\tau \, \in{} [-1,1]$.\footnote{The parameter $\tau$ corresponds to paramter t of \cite{eisert-1999-83}.} Positive $\tau$-values represent pure and mixed classical strategies, whereas negative $\tau$-values correspond to quantum strategies, where $\theta=0$ and $\varphi>0$. The whole strategy space is separated into four regions, namely the absolute classical region (CC: $\tau_A,\tau_B\geq0$), the absolute quantum region (QQ: $\tau_A,\tau_B<0$) and the two partially classical-quantum regions (CQ: $\tau_A\geq0 \wedge \tau_B<0$ and QC: $\tau_A<0 \wedge \tau_B\geq0$). It should be mentioned that within the $(\tau_A,\tau_B)$ representation the set of possible strategies $\{(\theta,\varphi) \mid \theta \in{} [0,\pi] \, , \,\, \varphi \in{} [0,\frac{\pi}{2}] \}$ is reduced to the following specific subset:
\begin{equation}
\underbrace{\{(\tau \, \pi,0) \mid \tau \in{} [0,1] \}}_{\hbox{classical region C}} \,\,\, \wedge \,\,\, \underbrace{\{(0,\tau \, \frac{\pi}{2}) \mid \tau \in{} [-1,0[ \}}_{\hbox{quantum region Q}} \quad .
\end{equation}

\subsection{Quantum Dove Strategies}

\begin{figure}[b]
\vspace*{-0.60cm}
\centerline{
\includegraphics[width=3.5in]{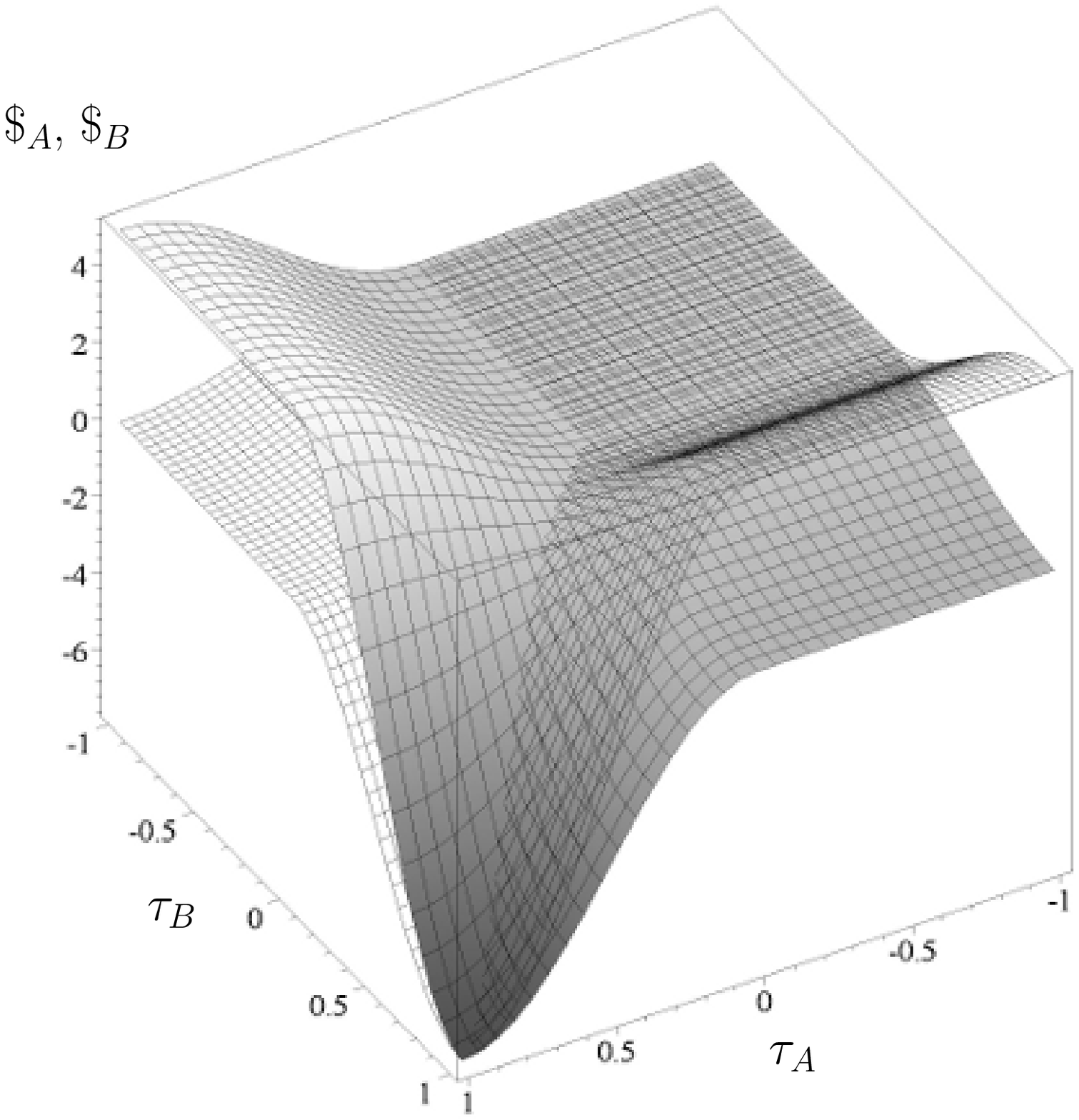}
}
\vspace*{-3.5cm}
\caption{Payoff surface of player A (solid) and player B (wired) as a function of their reduced strategies $\tau_A$ and $\tau_B$ within a non-entangled quantum game ($\gamma=0$) using the quantum dove strategy subset and the high risk parameter setting $P3$.}
\label{fig:highrisk0_a}
\end{figure}

As the $\theta$-value of the quantum region Q is fixed to zero, the possible quantum strategies can be understood as ''Quantum Dove'' strategies. In the following we will show results within this quantum dove strategy subset, where $\tau_A,\tau_B=1$ corresponds to strategy $H$, and $\tau_A,\tau_B=0$ corresponds to strategy $D$. All the results presented within this subsection where calculated using this quantum dove strategy subset. 

Diagram \ref{fig:highrisk0_a} illustrates the outcomes of the high risk game setting by visualising the payoff surfaces of investment banker A (solid surface) and investment banker B (wired surface) as a function of their strategies $\tau_A$ and $\tau_B$. In all of the presented three dimensional Figures (within this subsection) the absolute quantum region QQ is projected in the back, whereas the absolute classical region CC is projected to the front. Figure \ref{fig:highrisk0_a} shows the result where no strategic entanglement is present ($\gamma=0$). The diagram clearly exhibits that the non-entangled quantum game simply describes the classical version of the high risk Hawk-Dove game. For the case, that both players decide to play a quantum strategy ($\tau_A < 0 \wedge \tau_B < 0$) their payoff is equal to the case where both players choose the classical dove strategy $D$ ($ \$_A(D,D)=\$_A(\tau_A=0,\tau_B=0)=\frac{p_m}{2} $). The two classical non symmetric pure Nash equilibria ($(x=1,y=0) \hat{=} (H,D)$ and $(x=0,y=1) \hat{=} (D,H)$) correspond to the following $\tau$-values: $(H,D) \hat{=} (\tau_A=1,\tau_B=0)$ and $(D,H) \hat{=} (\tau_A=0,\tau_B=1)$. The ESS of the classical game (the mixed strategy Nash equilibrium $(x^*=\frac{p_m - 2 p_h}{p_m - p_h -d},y^*=\frac{p_m - 2 p_h}{p_m - p_h -d})$ is equal to the strategy point $(\tau_A^{*c},\tau_B^{*c})=(\frac{2}{\pi}\hbox{arccos}\!\left( \sqrt{1-\frac{p_m - 2 p_h}{p_m - p_h -d}} \right), \frac{2}{\pi}\hbox{arccos}\!\left(1- \sqrt{\frac{p_m - 2 p_h}{p_m - p_h -d}} \right) )$. At $(\tau_A^{*c},\tau_B^{*c})$ the partial derivatives $\frac{\partial \$_A}{\partial \tau_A}(\tau_A,\tau_B^{*c})$ and $\frac{\partial \$_B}{\partial \tau_B}(\tau_A^{*c},\tau_B)$ vanish for all possible strategy choices. 

\begin{figure}[t]
\vspace*{-0.30cm}
\centerline{
\includegraphics[width=3.4in]{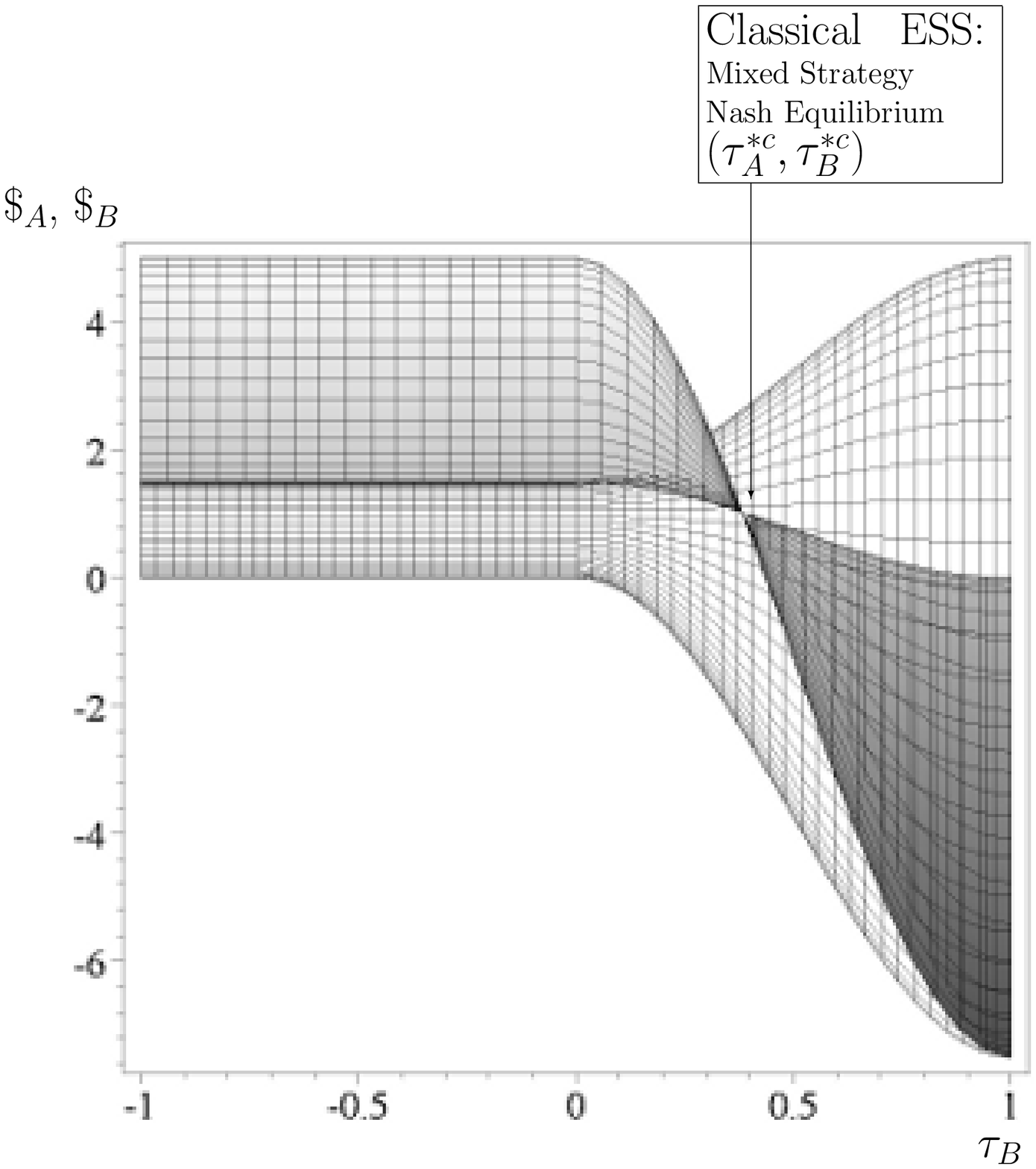}
}
\vspace*{-4.2cm}
\caption{Projection of the payoff surface of Figure \ref{fig:highrisk0_a} in direction of the $\tau_A$ axis.}
\label{fig:highrisk0_a_proj}
\end{figure}

\begin{eqnarray}
\left. \frac{\partial \$_A}{\partial \tau_A}(\tau_A,\tau_B) \right|_{\tau_B=\tau_B^{*c}} &=& 0 \,\,\, \forall \,\, \tau_A \in{} [-1,1]\\
\left. \frac{\partial \$_B}{\partial \tau_B}(\tau_A,\tau_B) \right|_{\tau_A=\tau_A^{*c}} &=& 0 \,\,\, \forall \,\, \tau_B \in{} [-1,1]\nonumber
\end{eqnarray}

\begin{figure}[b]
\vspace*{-1.00cm}
\centerline{
\includegraphics[width=3.0in]{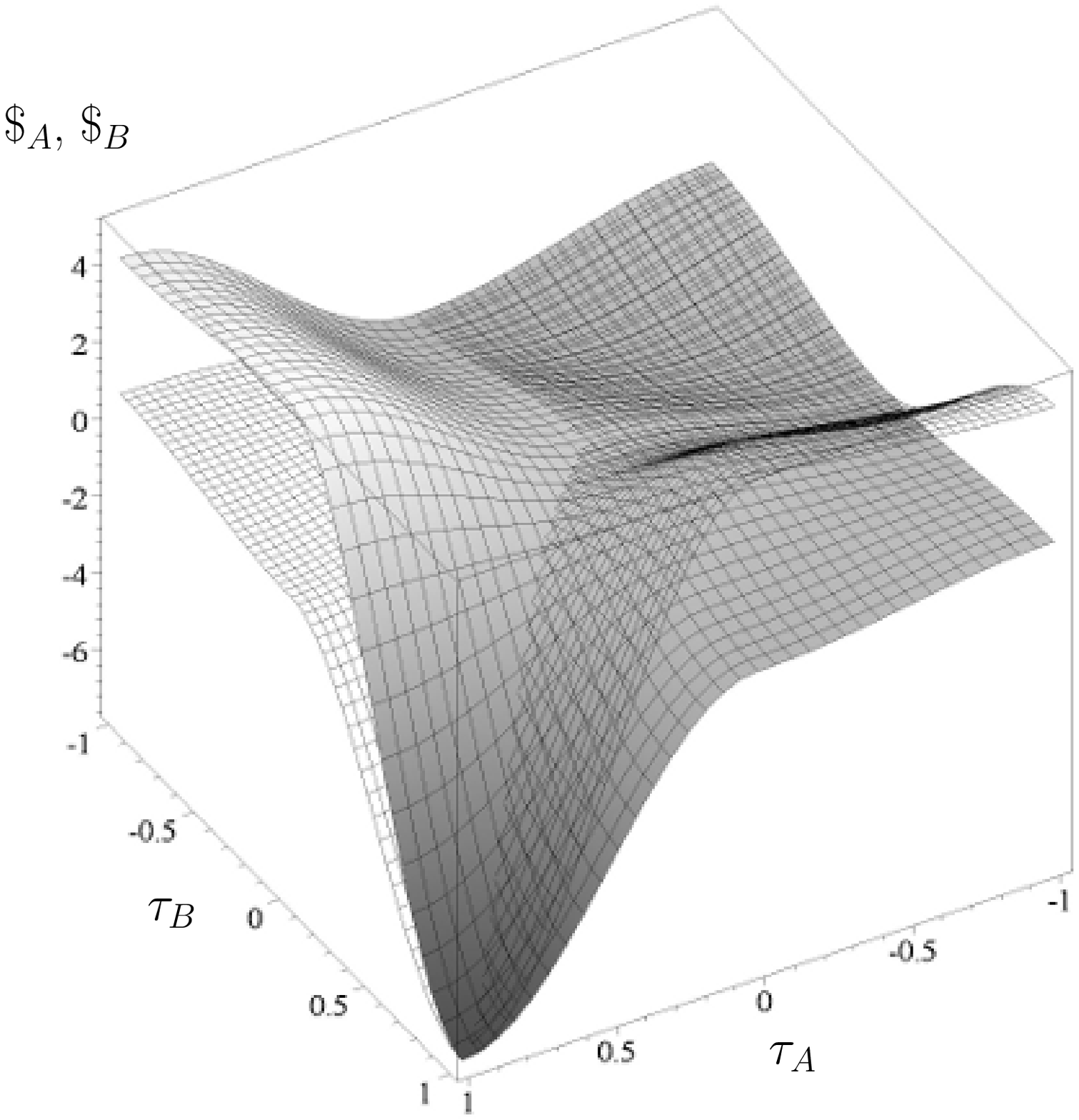}
}
\vspace*{-3.9cm}
\centerline{
\includegraphics[width=3.0in]{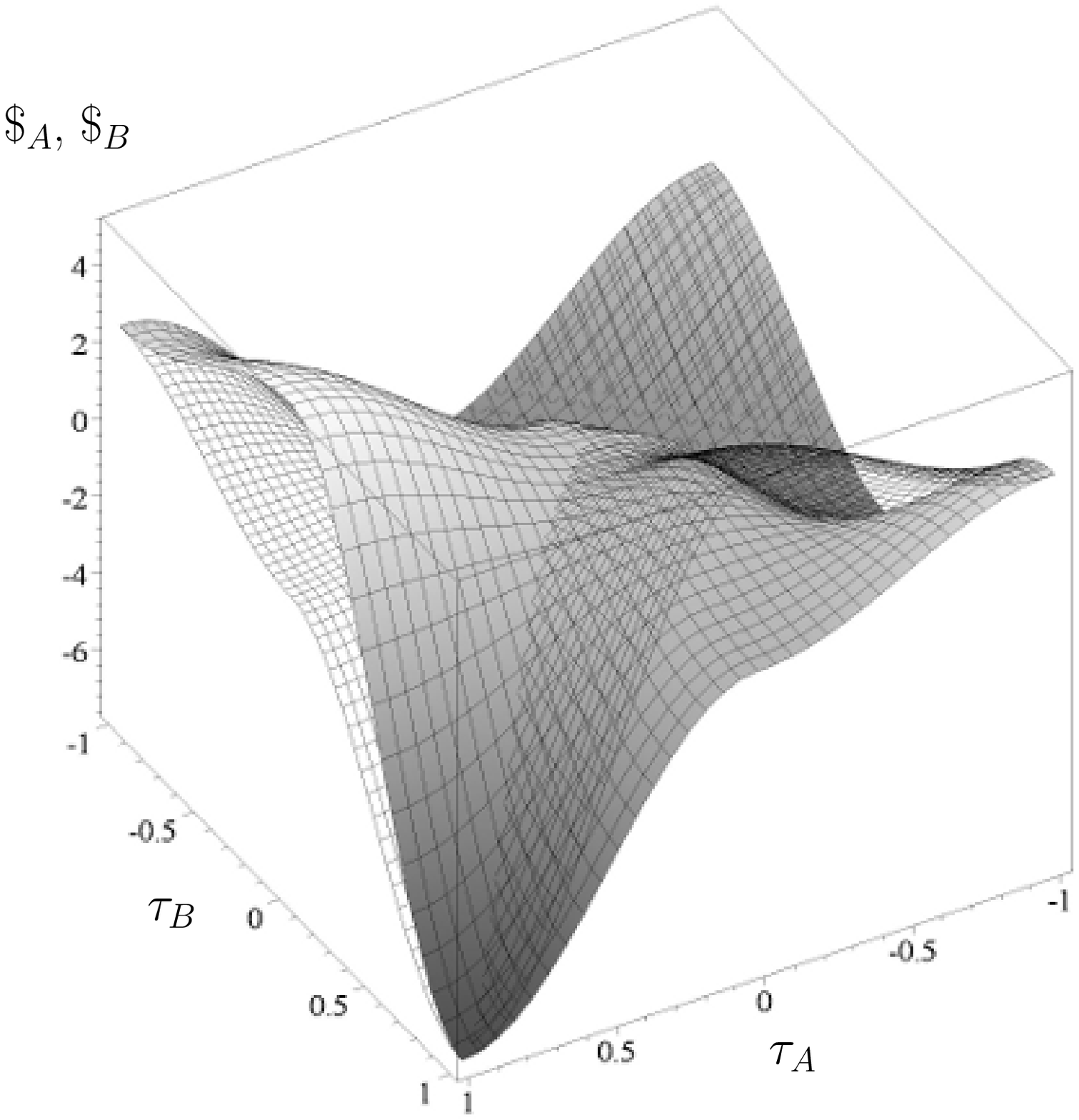}
}
\vspace*{-3.3cm}
\caption{Same description as Figure \ref{fig:highrisk0_a}, whereas the upper Figure is calculated within a sparsely entangled quantum game ($\gamma=\frac{\pi}{8}$) and the lower picture represents results within a medium entangled quantum game ($\gamma=\frac{\pi}{4}$).}
\label{fig:highriskPi4_a}
\end{figure}

This property of the classical ESS can be visualised by changing the projected viewpoint of the three dimensional surface. Figure \ref{fig:highrisk0_a_proj} shows again the payoffs of the investment bankers within the non-entangled quantum game, whereas the projection of the picture is now along the $\tau_A$-axis. As the partial derivative $\frac{\partial \$_A}{\partial \tau_A}(\tau_A,\tau_B^{*c})$ vanishes for all $\tau_A$-values, no gradient is observed at $\tau_B=\tau_B^{*c}$ and as a result the whole projected surface shrinks to one point (see Figure \ref{fig:highrisk0_a_proj}).

\begin{figure*}
\vspace*{-0.00cm}
\centerline{
\includegraphics[width=6in]{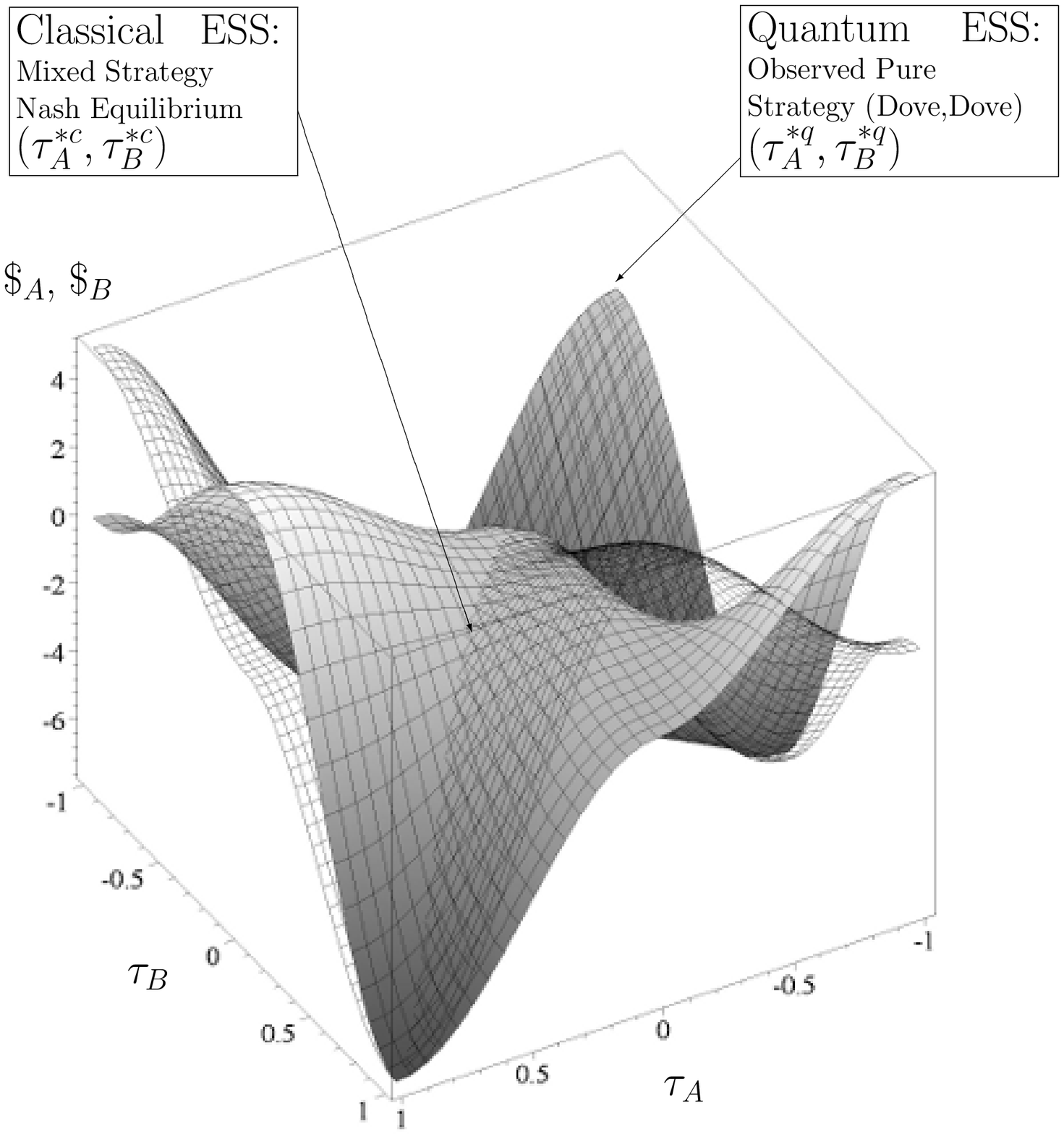}
}
\vspace*{-5.9cm}
\caption{Same description as Figure \ref{fig:highrisk0_a}, whereas the results where calculated within a maximally entangled quantum game ($\gamma=\frac{\pi}{2}$) using parameter set $P3$.}
\label{fig:highriskmax_a}
\end{figure*}

While the Figures \ref{fig:highrisk0_a} and \ref{fig:highrisk0_a_proj} visualise the non-entangled quantum game, Figure \ref{fig:highriskPi4_a} shows the payoff structure of a low (upper picture, $\gamma=\frac{\pi}{4}$) and medium (lower picture, $\gamma=\frac{\pi}{4}$) entangled high risk quantum game. The total classical region CC is equal to the non-entangled game (see Figure \ref{fig:highrisk0_a}), whereas in all other regions the shape of the payoff surfaces $\$_A$ and $\$_B$ has changed. The classical ESS and one of the asymmetric, pure strategy Nash equilibria ($(H,D) \hat{=} (\tau_A=1,\tau_B=0)$) still remain present in both diagrams, whereas the other pure strategy Nash equilibrium $(D,H) \hat{=} (\tau_A=0,\tau_B=1)$ 
disappears even for a tiny strength of entanglement. A further increase of entanglement will even change the structure of the existing ESS as a new, payoff dominant quantum ESS at ($\tau_A=-1,\tau_B=-1$) appears for $\gamma > 0.99$.

Figure \ref{fig:highriskmax_a} shows the payoff structure of the maximally entangled ($\gamma=\frac{\pi}{2}$) high risk quantum game within the quantum dove strategy subset. The classical ESS and one of the asymmetric, pure strategy Nash equilibria ($(H,D) \hat{=} (\tau_A=1,\tau_B=0)$) still remain present, while the pure classical Nash equilibrium $(D,H) \hat{=} (\tau_A=0,\tau_B=1)$ has vanished. Beside the remaining classical ESS $(\tau_A^{*c},\tau_B^{*c})$ a new quantum ESS ($(\tau_A^{*q},\tau_B^{*q})=(-1, -1)$) has been found for $\gamma > 0.99$. The point on the payoff surface, where both players choose the quantum ESS $\tau^{*q}$ is marked in Figure \ref{fig:highriskmax_a}. Which of these equilibria will be chosen by the whole population, is going to be addressed in section \ref{qeg} when the time evolution of quantum games is going to be discussed. As the payoff of this new quantum ESS ($\$_A(\tau_A^{*q},\tau_B^{*q})=\frac{p_m}{2}$) is higher than the payoff of the classical ESS ($\$_A(\tau_A^{*c},\tau_B^{*c}) \approx 1.02$), the fully entangled quantum players will likely asymptotically end within the new, payoff dominant quantum ESS. As the observable measurement of the strategy choice $(\tau_A^{*q},\tau_B^{*q})$ is the strategy set where both players play the dove strategy $D$ ($(\tau_A^{*q},\tau_B^{*q}) \hat{=} (D,D)$), fully entangled quantum players will likely end in a totaly dove strategy population ($x=0$).

\begin{figure}[t]
\vspace*{-0.4cm}
\centerline{
\includegraphics[width=3.5in]{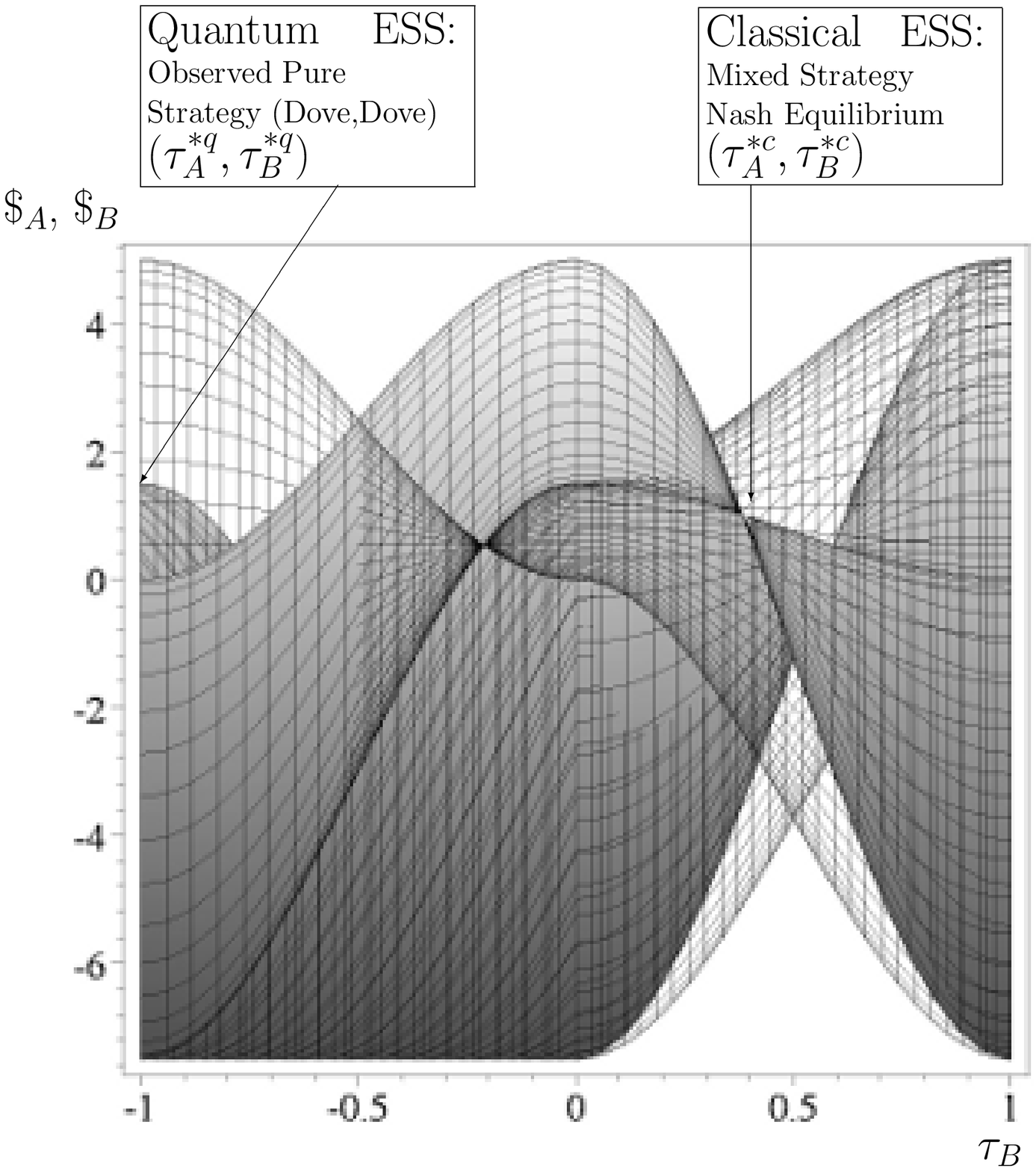}
}
\vspace*{-4.1cm}
\caption{Projection of the payoff surface \ref{fig:highriskmax_a} in direction of the $\tau_A$ axis.}
\label{fig:highriskmax_a_proj}
\end{figure}

To visualise the payoff values of the two ESSs more explicit, Figure \ref{fig:highriskmax_a_proj} projects the three dimensional surface of Figure \ref{fig:highriskmax_a} in direction of the $\tau_A$-axis. As the partial derivative of the classical ESS is only zero in the CC-region of the 3-dimensional plot, the whole surface does not shrink to one point as in Figure \ref{fig:highrisk0_a_proj}. 

\begin{figure}[b]
\vspace*{-1.00cm}
\centerline{
\includegraphics[width=2.8in]{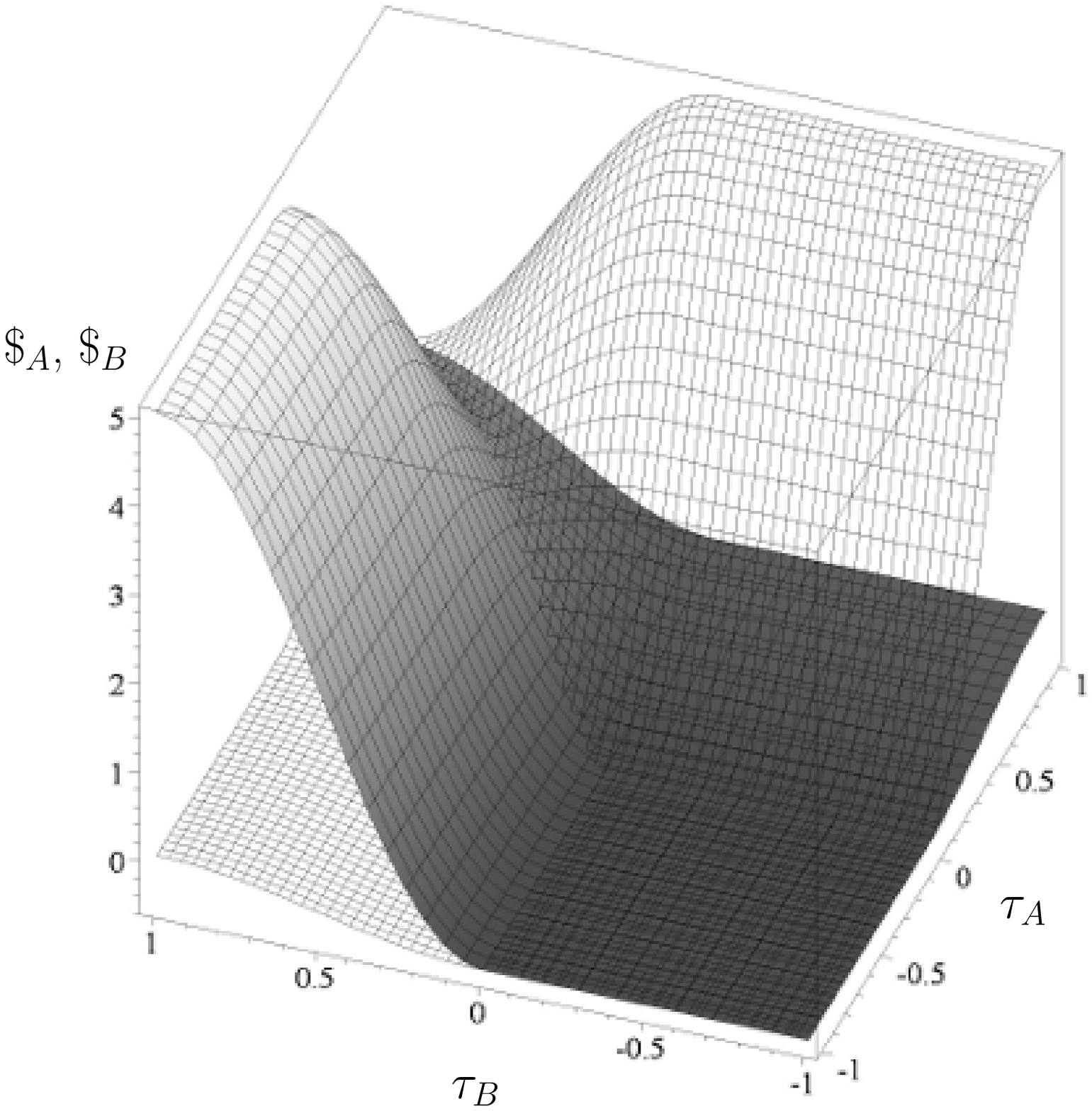}
}
\vspace*{-3.55cm}
\centerline{
\includegraphics[width=2.8in]{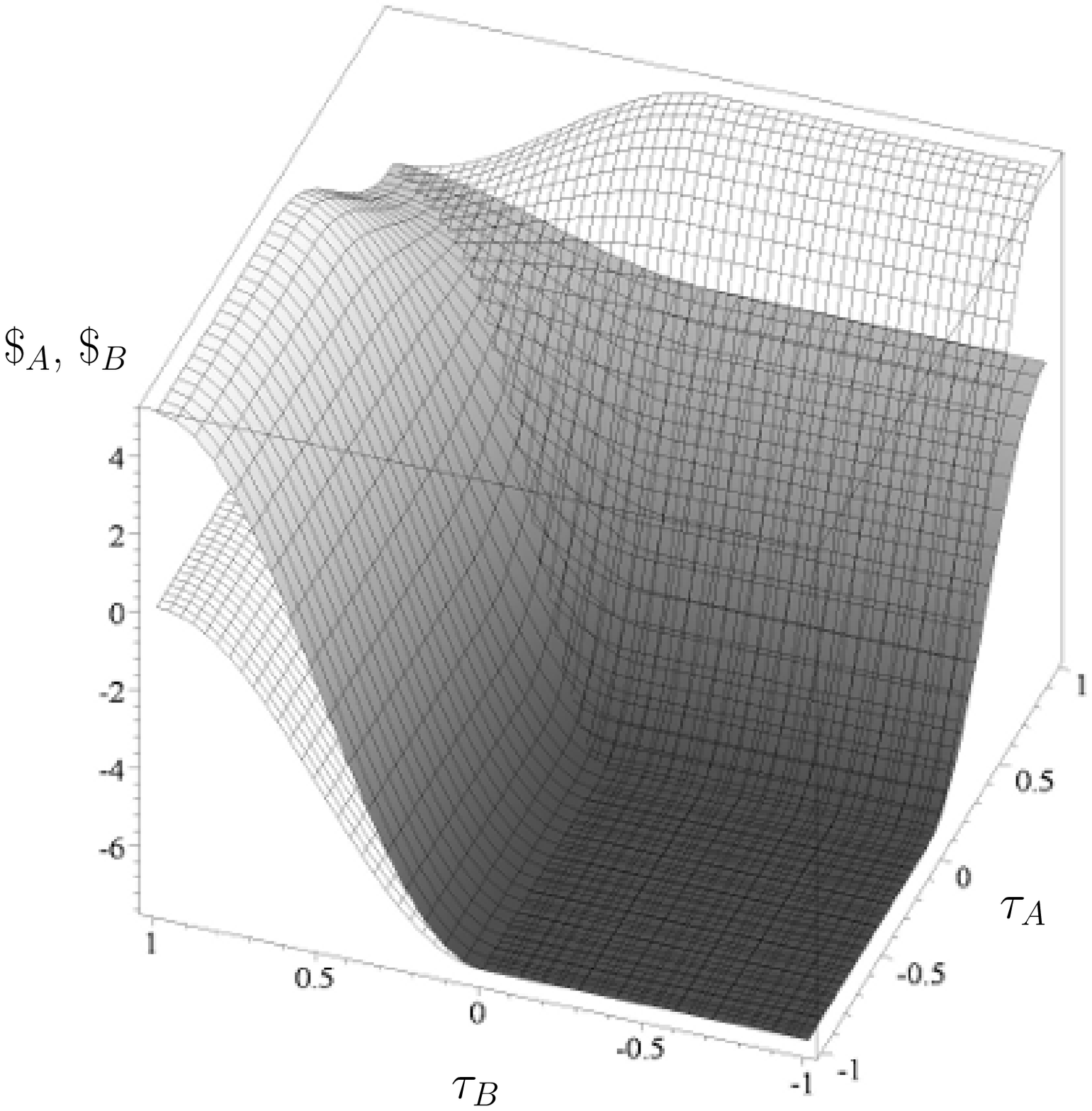}
}
\vspace*{-3.3cm}
\caption{Payoff surface of player A (solid) and player B (wired) as a function of their reduced strategies $\tau_A$ and $\tau_B$ within a non-entangled quantum game ($\gamma=0$) using the quantum-hawk strategy subset. The upper Figure depicts the results of the low risk parameter set $P1$, whereas the lower Figure shows the results of the high risk setting $P3$.}
\label{fig:low-highrisk0}
\end{figure}

\subsection{Quantum Hawk Strategies}
Within the previous subsection the set of possible strategies belong to the subset of quantum dove strategies whereas all the results presented within this subsection where calculated using the quantum hawk strategy subset. The corresponding  quantum game restricted on a quantum hawk strategy subset is constructed as follows: We redefine the four basis vectors of the Hilbert space ${\cal{H}}$ as the following classical game outcomes ($\left| HH \right>:=(1,0,0,0)$, $\left| HD \right>:=(0,-1,0,0)$, $\left| DH \right>:=(0,0,-1,0)$ and $\left| DD \right>:=(0,0,0,1)$). The setup of the quantum game begins with the choice of the initial state $\left| \Psi_0 \right>$, where we assume that both players are in the state $\left| H \right>$. The classical strategy H (Hawk) is now selected by appointing $\theta=0$ and $\varphi=0$ whereas the strategy D (Dove) is selected by choosing $\theta=\pi$ and $\varphi=0$. Finally, the state prior to detection is formulated as follows 
\begin{equation}
\left| \Psi_f \right> = \hat{\cal{J}}^\dagger \left( \hat{\cal{U}}_A \otimes \hat{\cal{U}}_B \right) \hat{\cal{J}}\, \left| HH \right> \quad ,
\end{equation}
where the entanglement operator $\widehat{\mathcal{J}}$ is formally given by $\widehat{\mathcal{J}} = e^{i \, \frac{\gamma}{2} (\widehat{H} \otimes \, \widehat{H})}$.

Within this quantum hawk strategy model $\tau_A,\tau_B=1$ corresponds to strategy $D$, and $\tau_A,\tau_B=0$ corresponds to strategy $H$. Negative $\tau$-values correspond again to quantum strategies, where $\theta=0$ and $\varphi>0$. As the $\theta$ value of the quantum region Q is fixed to zero which corresponds now to the classical hawk strategy, the possible quantum strategies can be understood as ''Quantum Hawk'' strategies. In the following we will show results within this quantum-hawk strategy subset.

\begin{figure*}
\vspace*{-0.00cm}
\centerline{
\includegraphics[width=6in]{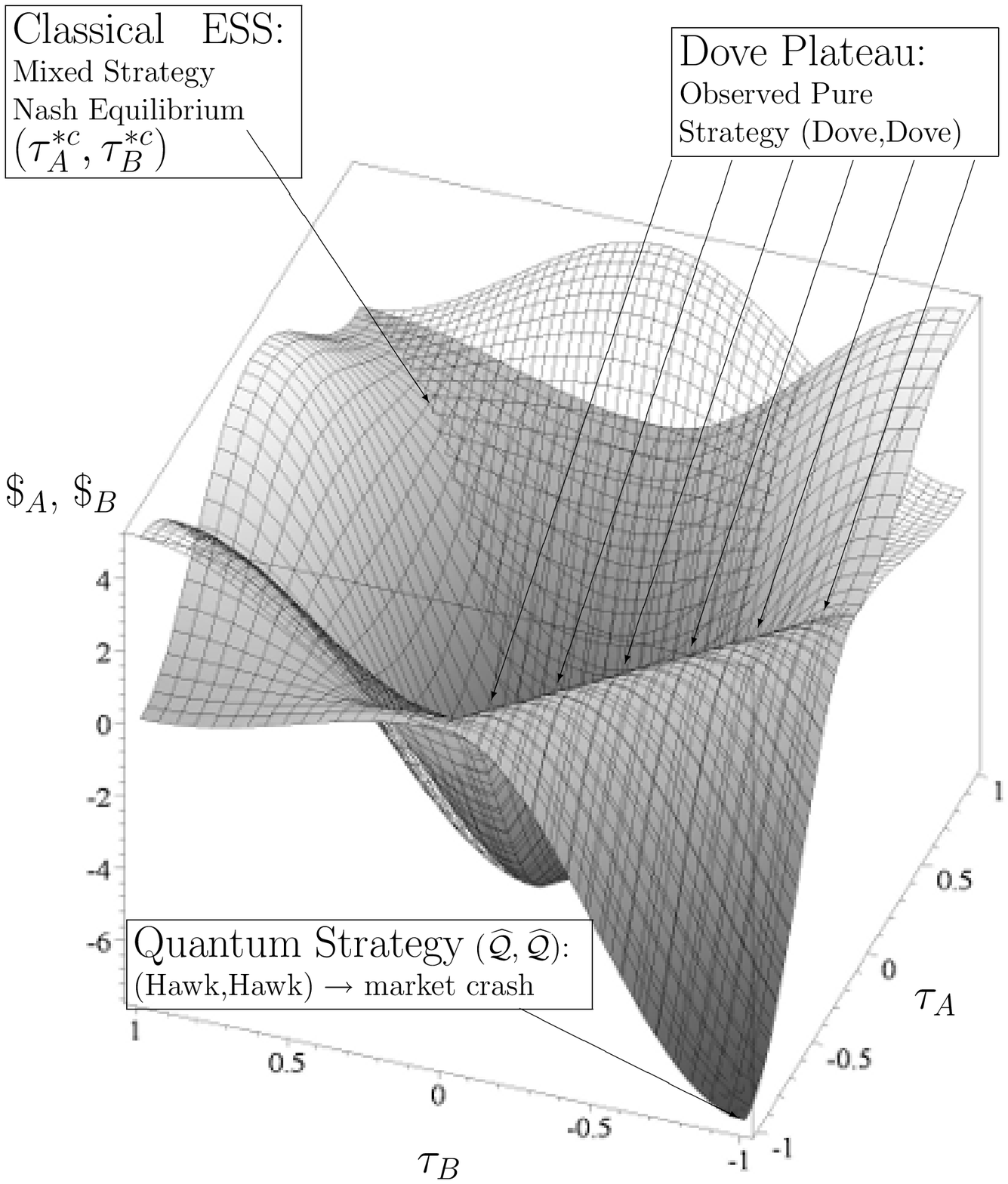}
}
\vspace*{-5.9cm}
\caption{Same description as Figure \ref{fig:low-highrisk0}, whereas the results where calculated within a maximally entangled quantum game ($\gamma=\frac{\pi}{2}$) using parameter set $P3$.}
\label{fig:highriskmaxa}
\end{figure*}

The two diagrams of Figure \ref{fig:low-highrisk0} illustrate the outcomes of the low and high risk game settings by visualising the payoff surfaces of investment banker A (solid surface) and investment banker B (wired surface) as a function of their strategies $\tau_A$ and $\tau_B$. Because of visual reasons, in all of the presented three dimensional Figures the absolute quantum region QQ is now projected in the front, whereas the absolute classical region is projected to the back. Figure \ref{fig:low-highrisk0} shows the result where no strategic entanglement is present ($\gamma=0$), where the upper Figure depicts the low risk parameter case P1 and the lower Figure shows the calculated results within the high risk parameter setting P3. Both diagrams clearly show that the non-entangled quantum game simply describes the classical versions of the low and high risk Hawk-Dove games. For the case, that both players decide to play a quantum strategy ($\tau_A < 0 \wedge \tau_B < 0$) their payoff is the games lowest payoff which is equal to the case, where both players choose the hawk strategy $H$ ($ \$_A(H,H)=\$_A(\tau_A=0,\tau_B=0)=\frac{p_h - d}{2} $). The two classical non symmetric pure Nash equilibria ($(x=1,y=0) \hat{=} (H,D)$ and $(x=0,y=1) \hat{=} (D,H)$) correspond now to the following $\tau$-values: $(H,D) \hat{=} (\tau_A=0,\tau_B=1)$ and $(D,H) \hat{=} (\tau_A=1,\tau_B=0)$. The ESS of the classical game (the mixed strategy Nash equilibrium) $(x^*=\frac{p_m - 2 p_h}{p_m - p_h -d},y^*=\frac{p_m - 2 p_h}{p_m - p_h -d})$ is equal to the strategy point $(\tau_A^{*c},\tau_B^{*c})=(\frac{2}{\pi}\hbox{arccos}\!\left( \sqrt{\frac{p_m - 2 p_h}{p_m - p_h -d}} \right), \frac{2}{\pi}\hbox{arccos}\!\left( \sqrt{\frac{p_m - 2 p_h}{p_m - p_h -d}} \right) )$. At $(\tau_A^{*c},\tau_B^{*c})$ the partial derivatives $\frac{\partial \$_A}{\partial \tau_A}(\tau_A,\tau_B^{*c})$ and $\frac{\partial \$_B}{\partial \tau_B}(\tau_A^{*c},\tau_B)$ vanish for all possible strategy choices.

\begin{figure}[t]
\vspace*{-1.00cm}
\centerline{
\includegraphics[width=3.5in]{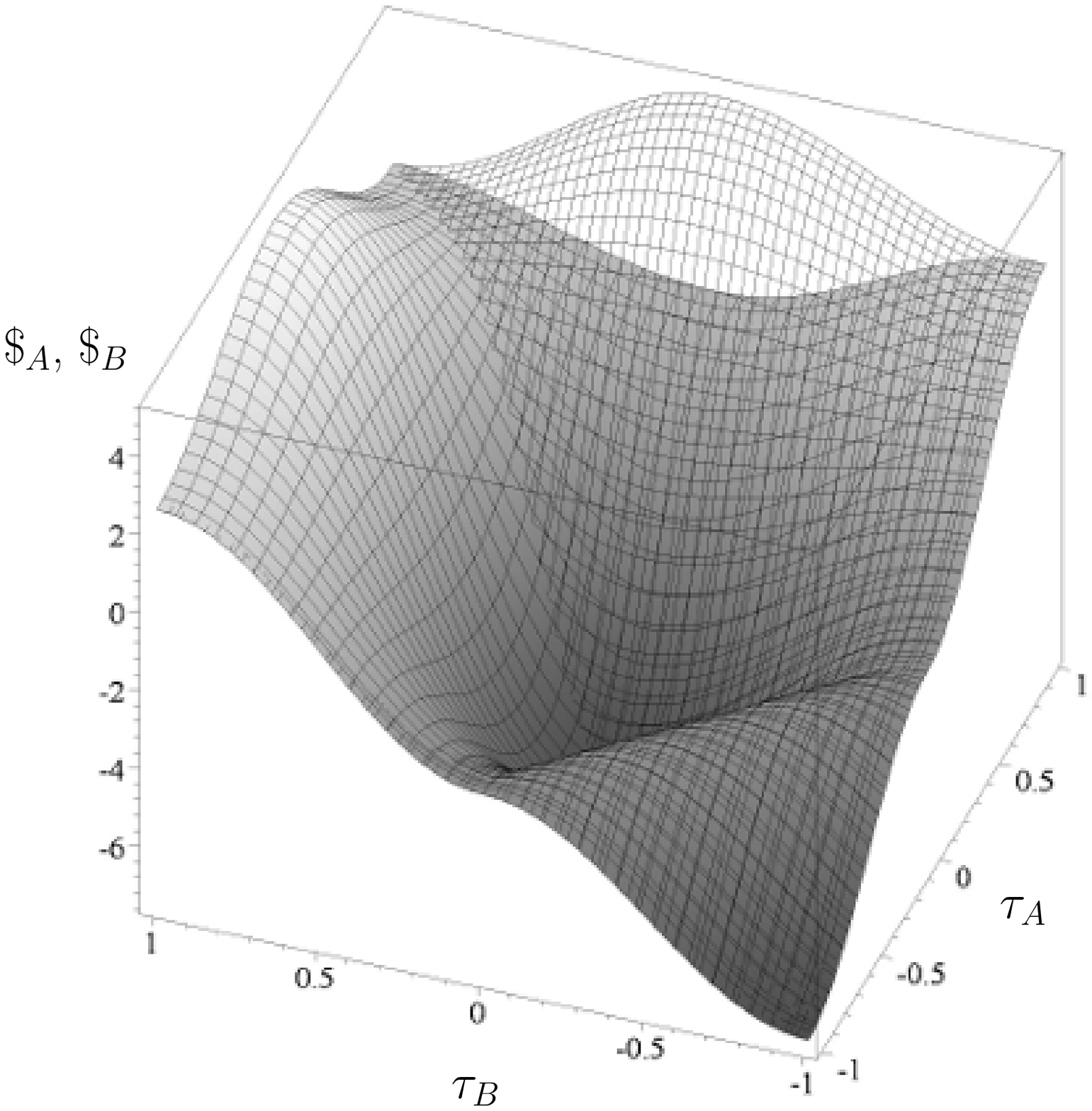}
}
\vspace*{-3.3cm}
\caption{Same description as Figure \ref{fig:low-highrisk0}, whereas the results where calculated within a medium entangled quantum game ($\gamma=\frac{\pi}{4}$) using parameter set $P3$.}
\label{fig:highrisk_Pi4}
\end{figure}

Figure \ref{fig:highriskmaxa} shows the payoff structure of the maximally entangled ($\gamma=\frac{\pi}{2}$) high risk quantum game. The classical ESS and one of the asymmetric, pure strategy Nash equilibria ($(H,D) \hat{=} (\tau_A=0,\tau_B=1)$) still remain present, while the pure classical Nash equilibrium $(D,H) \hat{=} (\tau_A=1,\tau_B=0)$ has vanished. Beside the remaining classical ESS $(\tau_A^{*c},\tau_B^{*c})$ a new quantum dove plateau has been found in the fully entangled quantum game. This new, relatively high payoff plateau is called the ''dove plateau'' because the observed measurement of a quantum strategy point at the top of it is the pure (D,D)-strategy and its payoff is $\frac{p_m}{2}$. It should be mentioned, that if both players decrease their $\tau$-value further than the $\tau$-value of the dove plateau their payoff extremly decrease. When the players choose the quantum strategy $\widehat{Q}$ $(\tau_A=-1,\tau_B=-1)$ in the maximally entangled high risk game, their payoff is equal to the lowest possible and their observed action is the hawk strategy ($(H,H)$). 

While the diagrams in Figure \ref{fig:low-highrisk0} visualise the non-entangled low and high risk quantum games, Figure \ref{fig:highrisk_Pi4} shows the payoff structure of the medium entangled ($\gamma=\frac{\pi}{4}$) high risk quantum game. The total classical region CC is equal to the non-entangled game (see Figure \ref{fig:low-highrisk0}, lower picture), whereas in all other regions the shape of the payoff surfaces $\$_A$ and $\$_B$ has changed. As the classical ESS and the asymmetric, pure strategy Nash equilibria ($(H,D) \hat{=} (\tau_A=0,\tau_B=1)$) and $(D,H) \hat{=} (\tau_A=1,\tau_B=0)$ still remain present, the outcome and the evolution of such a medium entangled quantum game will not be different from the classical situation. However, a further increase of the strength of entanglement will change the structure of the existing Nash equilibria. For $\gamma \geq 1.15$ the pure classical Nash equilibrium $(D,H) \hat{=} (\tau_A=1,\tau_B=0)$ disappears and for $\gamma \geq 1.34$ the dove plateau at the QQ-region (see Figure \ref{fig:highriskmaxa}) has a higher payoff than the payoff value of the classical ESS.

To summarise briefly the results of this section, we have shown on the one hand that within a highly entangled quantum version of the Hawk-Dove game a new, non aggressive ESS appears, but on the other hand the results indicate that when both players use a quantum hawk strategy and increase the quantum degree of their strategy ($\varphi$) beyond the dove-plateau, their payoff suddenly extremly decrease due to market destabilisation. 

\begin{figure}[b]
\vspace*{-1.00cm}
\centerline{
\includegraphics[width=3.5in]{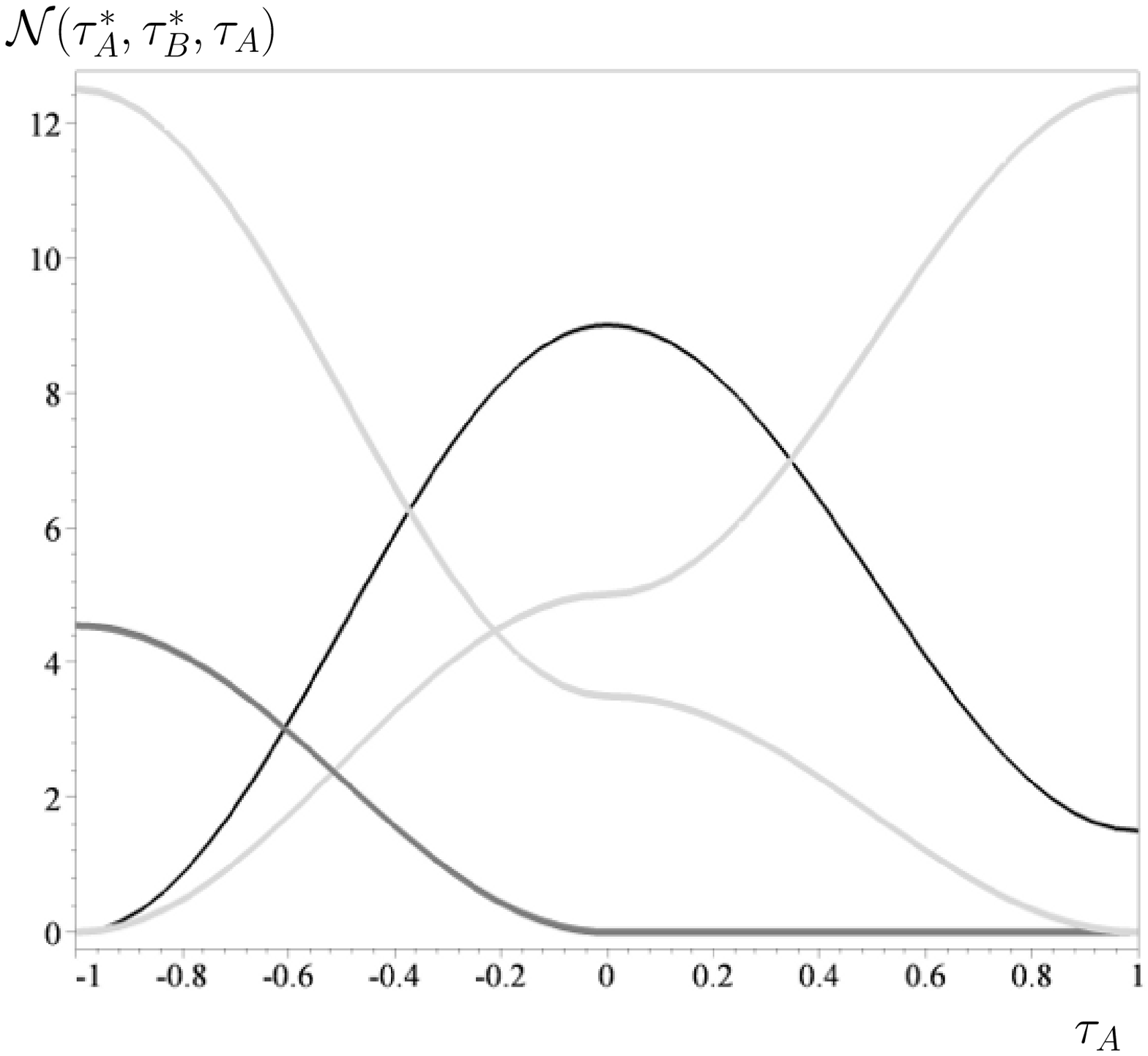}
}
\vspace*{-5.6cm}
\centerline{
\includegraphics[width=3.5in]{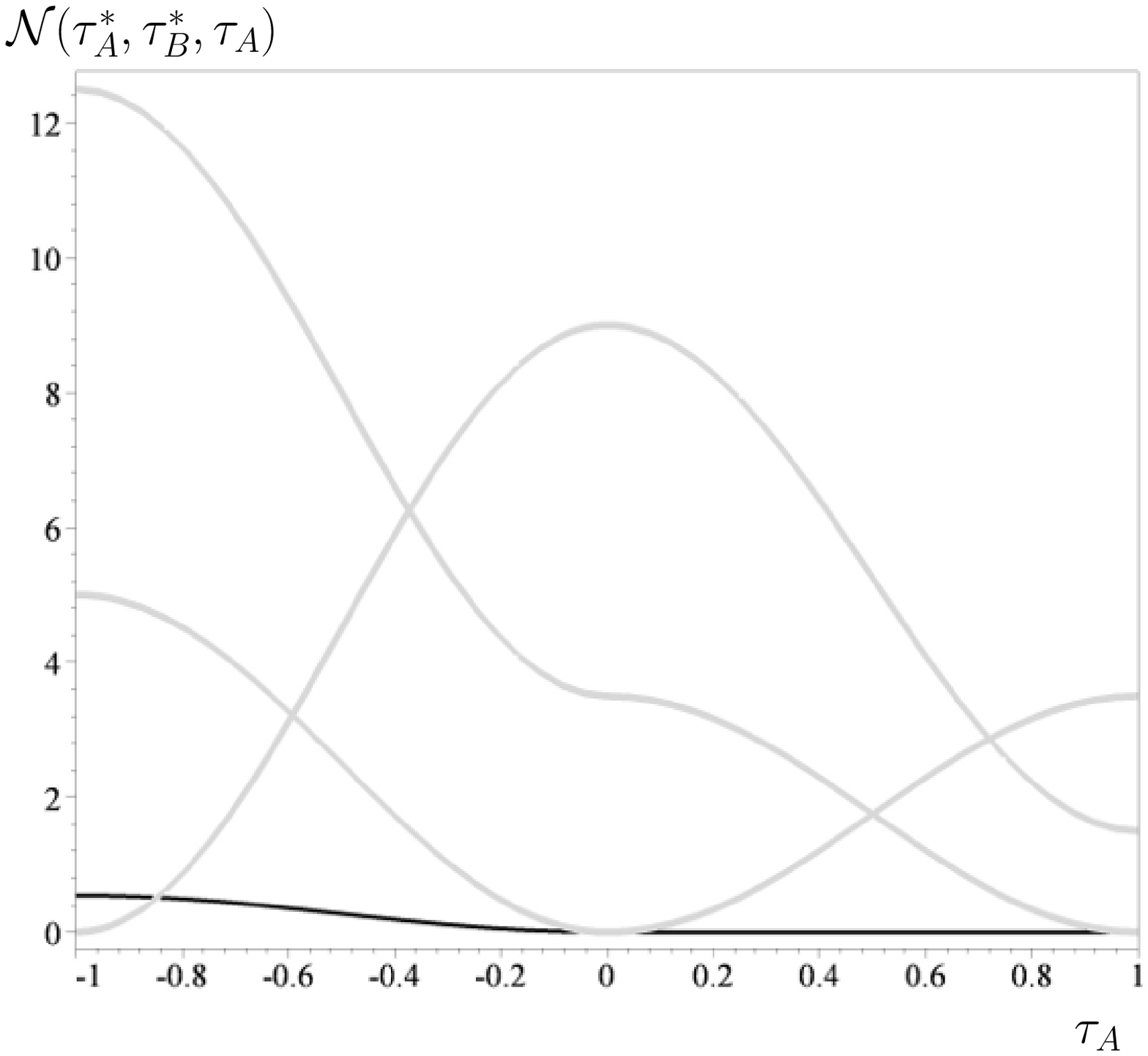}
}
\vspace*{-5.2cm}
\caption{${\cal N}(\tau_A^{*},\tau_B^{*},\tau_A)$ as a function of $\tau_A$ within the quantum dove strategy subset. The light grey curves visualise the non-symmetric Nash equilibria, whereas the dark grey and black curves depict ${\cal N}$ for the symmetric Nash equilibria. The upper picture shows the results calculated within the quantum dove strategy subset, whereas in the lower picture the calculation within the quantum hawk strategy subset are visualised.}
\label{fig:nq12_b}
\end{figure}

\section{The quantum evolutionary game of doves and hawks}\label{qeg}
In this section, the neccessary conditions ( a) and b), see section \ref{classical}) for the existence of ESSs are adopted to prove the existence of the new quantum ESS $\tau^{*q}$. The presented proof will be restricted to the maximally entangled game, but it can be shown, that it holds for any $\gamma > 0.99$. To illustrate that condition a) is fulfilled the pictures in Figure \ref{fig:nq12_b} depict the function ${\cal N}(\tau_A^{*},\tau_B^{*},\tau_A):=\$_A(\tau_A^{*},\tau_B^{*}) - \$_A(\tau_A,\tau_B^{*})$ versus $\tau_A$ for all symmetric and non-symmetric Nash equilibria. The upper diagram in Figure \ref{fig:nq12_b} shows the results within the quantum dove strategy subset, whereas the curves in the lower diagram of Figure \ref{fig:nq12_b} are calculated within the quantum hawk strategy subset. As all curves are always above zero they represent existent Nash equilibria. The function  ${\cal N}(\tau_A^{*},\tau_B^{*},\tau_A)$ for the two non-symmetric Nash equilibria in the upper picture ($\tau_A^{*}=-1,\tau_B^{*}=1$ and $\tau_A^{*}=1,\tau_B^{*}=0$) are visualised using light grey curves, the classical mixed strategy, symmetric Nash equilibrium $\tau_A^{*}=\tau_B^{*}=\tau^{*c}$ is illustrated with the dark grey curve, whereas the symmetric pure quantum Nash equilibrium $\tau_A^{*}=\tau_B^{*}=\tau^{*q}$ is shown by using a black curve. The Figure shows clearly, that within the quantum dove strategy subset two symmetric Nash equilibria and therefore two potential ESSs are present. Within the quantum hawk strategy subset (see lower diagram of Figure \ref{fig:nq12_b}) only one symmetric Nash equilibrium, the classical ESS is present (black curve). The other three, light grey curves represent the pure, non-symmetric Nash equilibria ($\tau_A^{*}=1,\tau_B^{*}=-1$, $\tau_A^{*}=-1,\tau_B^{*}=0$ and $\tau_A^{*}=0,\tau_B^{*}=1$). 

\begin{figure}[t]
\vspace*{-0.60cm}
\centerline{
\includegraphics[width=3.5in]{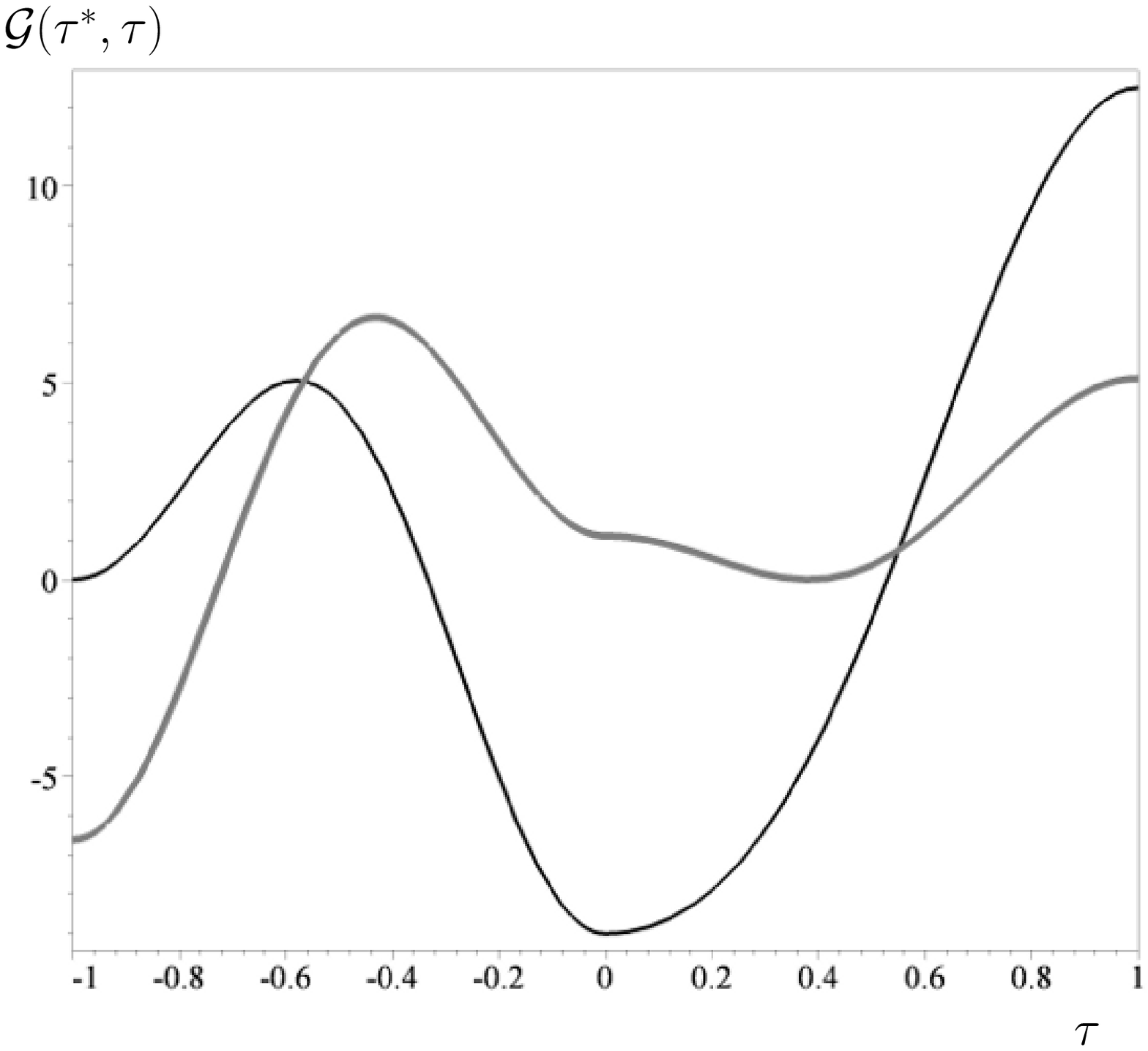}
}
\vspace*{-5.2cm}
\caption{${\cal G}(\tau^{*q},\tau)$ (black curve) and ${\cal G}(\tau^{*c},\tau)$ (grey curve) as a function of $\tau$ within the quantum dove strategy subset.}
\label{fig:nq3_b}
\end{figure}

To show that both of the symmetric Nash equilibria ($\tau^{*c}$ and $\tau^{*q}$) are ESSs, condition b) has additionally to be checked. Similar as in section \ref{classical} a function ${\cal G}(\tau^{*},\tau):=\$_A(\tau^{*},\tau) - \$_A(\tau,\tau)$ is defined, which has to be greater than zero for all strategies belonging to the set of best responses within the quantum dove strategy subset. At first we will varify, if the classical mixed strategy Nash equilibrium $\tau^{*c}$ remains an ESS for the maximally entangled quantum game. If player B chooses the strategy $\tau^{*c}$, the best response for player A are only strategies belonging to the CC-region, as the payoff of player A decreases within the QC-region (see Figure \ref{fig:highriskmax_a} in section \ref{quantum}). In the CC-region the derivative $\frac{\partial \$_A(\tau_A,\tau^{*c})}{\partial \tau_A}$ is equal to zero and as a result the set of best responses to the strategy $\tau^{*c}$ are all strategies belonging to the classical region ($r(\tau^{*c})=[0,1]$). The two curves in Figure \ref{fig:nq3_b} describe the functions ${\cal G}(\tau^{*c},\tau)$ (grey curve) and ${\cal G}(\tau^{*q},\tau)$ (black curve) as a function of the quantum strategy $\tau$ within the quantum dove strategy subset. As ${\cal G}(\tau^{*c},\tau)$ is greater than zero for all strategies $\tau \in [0,1] \ \tau^{*c}$, the classical mixed strategy Nash equilibrium, remains an ESS independent of the strength of entanglement. Secondly, we want to address the question whether the symmetric quantum Nash equilibrium $\tau^{*q}$ is indeed a new, additional ESS. If player B chooses the strategy $\tau^{*q}=-1$, the best response for player A is only again the strategy $\tau=-1$ and as a result condition b) is fulfilled, independently of the shape of the function ${\cal G}(\tau^{*q},\tau)$. Which of the ESS will be finally reached by the whole population will most likely depend on the initial conditions and on the underlying time dependent quantum dynamics. 

Quantum replicator dynamics (QRD), recently developed and discussed by E.G. Hidalgo \cite{esteban-2006,esteban-2008} (see also Toor et. al. \cite{toor-2001,toor-2001a}) was formulated within the density matrix approach of quantum game theory \cite{marinatto-2000-272}. QRD employs the von Neumann equation, which describes how a quantum density operator evolves in time. In order to reveal that the von Neumann equation is simply a quantum amplification of classical replicator dynamics (see equation \ref{Glei:Repro}), Hidalgo had reformulated equation \ref{Glei:Repro} to a matrix equation. Constraining to only two possible pure strategies equation \ref{Glei:Repro} can be formulated as follows \cite{esteban-2006}:
\begin{eqnarray}
&\frac{d}{dt} \widehat{X} = \left[ \widehat{\Lambda} , \widehat{X} \right]& \label{Glei:Repro_matrix}\\
&\widehat{X}:=
\left(
\begin{array}[c]{cc}
x_1&\sqrt{x_1 \, x_2}\\
\sqrt{x_2 \, x_1}&x_2
\end{array}
\right) \, \,\,\,
\widehat{\Lambda}:=
\left(
\begin{array}[c]{cc}
\Lambda_{11}&\Lambda_{12}\\
\Lambda_{21}&\Lambda_{22}
\end{array}
\right)& \nonumber\\
&\Lambda_{ij}:= \frac{1}{2} \sum_{k=1}^{n=2} \left(  \$_{ik} \, x_k \, \sqrt{x_i \, x_j} - 
\sqrt{x_j \, x_i} \, \$_{jk} \, x_k \right)&\nonumber
\end{eqnarray}
, where the matrix $\widehat{X}$ is an amplification of the population vector $\vec{x}=(x_1, x_2)$, $\left[ \widehat{a} , \widehat{b} \right]:= \widehat{a} \widehat{b} - \widehat{b} \widehat{a}$ is the commutator of the two matrices $\widehat{a}$ and $\widehat{b}$ and $\widehat{\Lambda}$ is a payoff dependent ($2 \times 2$)-matrix. The quantum amplification of classical replicator dynamics is realised by substituting of $\widehat{X}$ to the density matrix $\widehat{\rho}$ and  $\widehat{\Lambda}$ as the Hamilton Operator $\widehat{E}$ of the quantum system
\begin{equation}
E_{ij}:= \frac{1}{2} \sum_{k=1}^{n=2} \left(  \$_{ik} \, \rho_{kk} \, \rho_{ij} - 
\rho_{ji} \, \$_{jk} \, \rho_{kk} \right) \quad.
\end{equation}
Quantum replicator dynamics as an extension of equation \ref{Glei:Repro} and \ref{Glei:Repro_matrix} is described with the von Neumann equation 
\begin{equation}
\frac{d}{dt} \widehat{\rho} = \sigma \, \left[ \widehat{E} , \widehat{\rho} \right] \quad, \label{Neumann}
\end{equation}
where $\sigma$ is a certain quantisation constant.\footnote{The von Neumann equation usually describes how a quantum density operator $\widehat{\rho}$ evolves in time, where $\sigma:=\frac{1}{i \hbar}$. In quantum replicator dynamics $\sigma$ shall be deemed to be a certain constant.} The numerical simulation of equation \ref{Neumann} and therefore the time evolution of the Hawk-Dove quantum game is under construction and will be adressed in a seperate article.

\section{Interpretation and consequences}\label{conclusions}
With respect to the analysed strategy space the previous study provides the following results: Regarding the combination of classical and dove quantum strategies an additional ESS occurred in case of a high degree of entanglement. With respect to the combination of classical and hawk quantum strategies no additional ESS could be observed. Instead we found a dove-plateau and a steep reduction of payoffs behind this plateau. 

With respect to the financial crisis especially the first finding is of interest. Obviously, the players did not behave in the way that a highly entangled quantum game would suggest, as a certain proportion of bankers continued their risk-seeking, aggressive behaviour resulting in a market crash. Hence, in this situation no or only relatively little entanglement of the individuals’ decisions existed, which induced them to follow the classical ESS. Consequently, in order to induce the wished for behaviour – strict selection of non aggressive strategies – one has to introduce a high degree of entanglement into this economic situation. 

So far, in literature entanglement has been discussed from a more physical point of view. However, in order to derive consequences from the obtained results we want to propose one possibility to interpret it in an economic context. In this paper, entanglement has been termed a conjoint, psychological contract between the members of an economic population aligning their strategies. However, this contract is not the result of conscious negotiations but of general socio-economic factors influencing the agents simultaneously. These factors comprise moral standards, values, legal rules, joint experiences, a similar educational background etc. All these factors can drive the decision processes of different individuals into the same direction without the necessity that the individuals have to communicate to each other. The objective existence of these background factors can vary, which is reflected by the degree of the entanglement parameter . 

As the results show, if a certain degree of entanglement is surpassed, a new ESS appears in the space of quantum dove strategies. Hence, then it is more rational for individuals to choose a quantum dove strategy instead of a classical mixed strategy. At this point also the notion of quantum strategies has to be interpreted in a more economic sense. We propose the following point of view: While the degree of the parameter  exhibits the objective entanglement, i.e. the objectively observable influence of different socio-economic factors, the parameter $\varphi$, which has to be chosen by the agents when selecting a strategy, reflects the degree to which an agent actually considers theses factors during the decision process. The higher the degree of this parameter is, the more attention the agent pays to these factors. 

The results of the analysis point to the fact that if the degree of objective entanglement surpasses a certain threshold, it is rational for an agent to pay a lot of attention to the socio-economic factors and – in the context of the analysed situation – to play a non aggressive strategy. In contrast, as long as entanglement is below this threshold, it is more rational for the agents to ignore these factors and play a classical evolutionary game. In sum, the degree of objective entanglement also will determine the degree of subjective attention paid by the agents toward this entanglement.   

This is exactly the starting point for leading the agents’ decisions into the wished for direction: So far, being greedy and aggressive was either not seen as negative or even accepted as an adequate behavioural strategy in the community of investment bankers. However, this behavioural code can be modified through different measures: One important instrument is education. By teaching adequate values and behavioural rules in the institutions that train future investment bankers or market participants in general the value basis of these individuals can be changed in a way that favours less aggressive behaviour. Moreover, the strong disapproval of aggressive behaviour from the general public outside this community can introduce pressure to align one’s behaviour according to a less aggressive way. Furthermore, investment bankers were paid through bonus systems that rewarded aggressive and punished non aggressive actions. Hence, under these incentive schemes a reduction of aggressive behaviour was impossible, since they also fostered the feeling that being aggressive was a positively valued behavioural strategy. In order to change this connotation legal structures – as another part of the socio-economic context – have to be modified in a way that prevents this kind of payment systems. 

\section{Summary}\label{sec:sum}
The last financial and economical crisis demonstrated the dysfunctional long-term effects 
of aggressive behaviour in financial markets. Starting from this observation, this paper 
picked up a result of evolutionary game theory which states that under the condition of 
strategic dependence a certain degree of aggressive behaviour remains within a given 
population of agents and asked how one could change the ''rules of the game'' in a way that 
prevents the occurrence of any aggressive behaviour and thereby also the danger of market 
crashes. In order to answer this question we extended the in literature well-known 
evolutionary Hawk-Dove game by a quantum approach and analysed three scenarios in depth.

The resulting study exhibited that dependent on entanglement, also evolutionary stable 
strategies can emerge, which are not predicted by classical evolutionary game theory and 
where the total economic population uses a non aggressive quantum strategy. Hence, the 
obtained outcomes point into a direction, how the mentioned ''rules of the game'' could be 
changed to prevent future crashes.

In order to make this mathematical result actually usable in an economic context, we 
additionally provided an interpretation of the outcomes of our study in the context of 
economic situations: We transformed the more physical notions {\it entanglement} and {\it quantum 
strategies} into concepts of the analysed economic situation. We interpret entanglement 
as the objective influence of socio-economic context factors, while in this context the 
application of quantum strategies exhibits the degree to which decision makers 
incorporate these factors into their decisions. Under this premise, our results point to 
the importance of deliberately changing existing socio-economic context factors and 
thereby influencing market participants. In this context, we explicitly mentioned the 
provision of a value basis that prevents aggressive behaviour through educational 
measures, the strengthening of disapproval regarding aggressive behaviour in an economic 
context through the general public and the change of the legal basis for the provision of 
variable payment systems.

\section*{Acknowledgments}
M.H. wants to thank John Forbes Nash Jr. for the inspiring discussion during the Third Congress of the Game Theory Society (Games 2008). The conversation initiated and motivated the author to broaden his work to an evolutionary context. J.K. wants to thank Carsten Heineke for helpful comments especially regarding the economic interpretation.
\bibliographystyle{abbrv}

\begin{thebibliography}{10}

\bibitem{book-axelrod}
R.~Axelrod.
\newblock {\em The complexity of cooperation: Agent-based models of conflict
  and cooperation}.
\newblock The Princeton University Press, Princeton, 1997.

\bibitem{benjamin-2001-64}
S.~C. Benjamin and P.~M. Hayden.
\newblock Multi-player quantum games.
\newblock {\em Physical Review A}, 64:030301, 2001.
\newblock
  \href{http://www.citebase.org/abstract?id=oai:arXiv.org:quant-ph/0007038}{qu%
ant-ph/0007038}.

\bibitem{book-dopfer}
K.~Dopfer.
\newblock {\em Evolutionary economics: program and scope}.
\newblock Kluwer Academic Publishers, Boston, 2001.

\bibitem{dosi-94}
G.~Dosi and R.~R. Nelson.
\newblock {An introduction to evolutionary theories in economics}.
\newblock {\em Journal of Evolutionary Economics}, 4:153--172, 1994.

\bibitem{eisert-1999-83}
J.~Eisert, M.~Wilkens, and M.~Lewenstein.
\newblock Quantum games and quantum strategies.
\newblock {\em Physical Review Letters}, 83:3077, 1999.
\newblock
  \href{http://www.citebase.org/abstract?id=oai:arXiv.org:quant-ph/9806088}{qu%
ant-ph/9806088}.

\bibitem{friedman-98}
D.~Friedman.
\newblock {On economic applications of evolutionary game theory}.
\newblock {\em Journal of Evolutionary Economics}, 8:15--43, 1998.

\bibitem{esteban-2006}
E.~Guevara.
\newblock {Quantum Replicator Dynamics}.
\newblock {\em Physica A}, 369/2:393--407, 2006.
\newblock
  \href{http://arxiv.org/abs/quant-ph/0510238}{arXiv:quant-ph/0510238v7}.

\bibitem{esteban-2008}
E.~Guevara.
\newblock {Quantum Games and the Relationships between Quantum Mechanics and
  Game Theory}.
\newblock 2008.
\newblock \href{http://arxiv.org/abs/0803.0292}{arXiv:0803.0292v1}.

\bibitem{hanauske-2006}
M.~Hanauske, S.~Bernius, and B.~Dugall.
\newblock {Quantum Game Theory and Open Access Publishing}.
\newblock {\em Physica A}, 382(2):650--664, 2007.
\newblock
  \href{http://www.citebase.org/abstract?id=oai:arXiv.org:physics/0612234}{phy%
sics/0612234}.

\bibitem{book-hodgson}
G.~M. Hodgson.
\newblock {\em Economics and Evolution: Bringing Life back into Economics}.
\newblock The University of Michigan Press, Ann Arbor, 1993.

\bibitem{toor-2001}
A.~Iqbal and A.~H. Toor.
\newblock {Equilibria of Replicator Dynamics in Quantum Games}.
\newblock 2001.
\newblock
  \href{http://arxiv.org/abs/quant-ph/0106135}{arXiv:quant-ph/0106135v2}.

\bibitem{book_kiefer}
E.~Joos, H.~D. Zeh, C.~Kiefer, D.~Giulini, J.~Kupsch, and I.~O. Stamatescu.
\newblock {\em {Decoherence and the Appearance of a Classical World in Quantum
  Theory}}.
\newblock Springer, 2003.

\bibitem{book-malthus}
T.~R. Malthus.
\newblock {\em An essay on the principle population as it aects the future
  improvement of society with remarks on the speculations of Mr. Godwin, M.
  Condorcet and other writers}.
\newblock Johnson, London, 1798.

\bibitem{marinatto-2000-272}
L.~Marinatto and T.~Weber.
\newblock A quantum approach to static games of complete information.
\newblock {\em Physics Letters A}, 272:291, 2000.
\newblock
  \href{http://www.citebase.org/abstract?id=oai:arXiv.org:quant-ph/0004081}{qu%
ant-ph/0004081}.

\bibitem{meyer-1999-82}
D.~A. Meyer.
\newblock Quantum strategies.
\newblock {\em Physical Review Letters}, 82:1052, 1999.
\newblock
  \href{http://www.citebase.org/abstract?id=oai:arXiv.org:quant-ph/9804010}{qu%
ant-ph/9804010}.

\bibitem{miekisz-2007}
J.~Miekisz.
\newblock {Evolutionary game theory and population dynamics}.
\newblock 2007.
\newblock \href{http://arxiv.org/abs/q-bio/0703062}{q-bio/0703062}.

\bibitem{toor-2001a}
A.~Nawaz and A.H.Toor.
\newblock {Evolutionarily Stable Strategies in Quantum Hawk-Dove Game}.
\newblock 2001.
\newblock
  \href{http://arxiv.org/abs/quant-ph/0108075}{arXiv:quant-ph/0108075v2}.

\bibitem{book-osborne}
M.~J. Osborne and A.~Rubinstein.
\newblock {\em A course in game theory.}
\newblock MIT Press, Cambridge, Massachusetts, 1994.

\bibitem{piotrowski-2002-312}
E.~W. Piotrowski and J.~Sladkowski.
\newblock Quantum {M}arket {G}ames.
\newblock {\em Physica A}, 312:208, 2002.
\newblock
  \href{http://www.citebase.org/abstract?id=oai:arXiv.org:quant-ph/0104006}{qu%
ant-ph/0104006}.

\bibitem{Sally.2001}
D.~Sally.
\newblock On sympathy and games.
\newblock {\em Journal of Economic Behavior and Organization}, 44(1):1--30,
  2001.

\bibitem{book_schlee_gametheory}
W.~Schlee.
\newblock {\em {Einf\"uhrung in die Spieltheorie}}.
\newblock Vieweg, 2004.

\bibitem{schlosshauer-2004}
M.~Schlosshauer.
\newblock {Decoherence, the measurement problem, and interpretations of quantum
  mechanics}.
\newblock {\em Rev. Mod. Phys.}, 76:1267--1305, 2004.
\newblock
  \href{http://arxiv.org/abs/quant-ph/0312059}{arXiv:quant-ph/0312059v4}.

\bibitem{book-smith1776}
A.~Smith.
\newblock {\em An inquiry into the nature and causes of the wealth of nations}.
\newblock Strahan and Cadell, London. Reprint in Oxford University Press,
  Oxford 1993, 1776.

\bibitem{smith-1972}
J.~M. Smith.
\newblock {Game theory and the evolution of fighting}.
\newblock pages 8--28, 1972.
\newblock {\it On Evolution}, Edinburgh University Press.

\bibitem{book_smith}
J.~M. Smith.
\newblock {\em Evolution and the Theory of Games}.
\newblock Cambridge University Press, 1982.

\bibitem{smith-1986}
J.~M. Smith.
\newblock {Evolutionary game theory}.
\newblock {\em Physica D}, 22:43--49, 1986.

\end{thebibliography}

\end{document}